\documentclass[aps,prd,nofootinbib,superscriptaddress]{revtex4-1} 
\usepackage[english]{babel}
\usepackage{amsmath} % Useful for equation alignment
\usepackage{verbatim} % Allows to have comments over multiple lines
\usepackage{mathrsfs}
\usepackage{amsfonts}
\usepackage[latin1]{inputenc}
\usepackage{hyperref}
\usepackage{graphicx}
\usepackage{subfig}

\newcommand{\pyRI}{\hat{\pi}^{R,I}}

\newcommand{\fluc}[1]{(\Delta #1)^2}      % square of fluctuations
 
\newcommand{\cre}{\hat{a}^{\dagger}}    % creation

\newcommand{\ann}{\hat{a}}            % and annihilation operators

\newcommand{\tc}{\eta_{\vec{k}}^c}
\newcommand{\beq}{\begin{equation}}
\newcommand{\eeq}{\end{equation}}
\newcommand{\nk}{\vec{k}}
\newcommand{\dphi}{\delta \phi}
\newcommand{\x}{\vec{x}}

\newcommand{\bra}{\langle}
\newcommand{\ket}{\rangle}
\newcommand{\bea}{\begin{eqnarray}}
\newcommand{\eea}{\end{eqnarray}}

\newcommand{\almsix}{\overline{a_{l_1 m_1} a_{l_2 m_2} a_{l_3 m_3} a_{l_1 m_4}^\star a_{l_2 m_5}^\star a_{l_3 m_6}^\star}}

\begin{document}

\author{Gabriel Le\'on}
\email{gleon@df.uba.ar}
\affiliation{Departamento de F\'isica, Facultad de Ciencias Exactas y Naturales, Universidad de Buenos Aires, Ciudad Universitaria, Buenos Aires 1428, Argentina}

\author{Daniel Sudarsky}
\email{sudarsky@nucleares.unam.mx}
\affiliation{Instituto de Ciencias Nucleares, Universidad Nacional Aut\'onoma de M\'exico, M\'exico D.F. 04510, M\'exico}

%\title{Origin of structure: Primordial Bispectrum without non-Gaussianities}
\title{Origin of structure:  Statistical characterization of  the  primordial    density fluctuations   and the collapse of the wave function}

\begin{abstract}
The statistical properties  of the  primordial density  perturbations  has been considered in the past decade as a powerful probe of the physical processes taking place  in the early universe. Within the inflationary paradigm, the properties of the bispectrum are one  of the keys  that serves to discriminate among competing scenarios concerning the details  of the origin of cosmological perturbations.  However,  all of the scenarios, based on the conventional approach to  the so-called  ``quantum-to-classical transition'' during  inflation, lack the ability to point out the precise  physical mechanism responsible for generating the inhomogeneity and anisotropy of our universe starting  from and exactly homogeneous and isotropic vacuum state associated with the early inflationary regime. In past works, we have shown that the proposals  involving a spontaneous  dynamical reduction of the quantum state provide plausible explanations for the birth of said  primordial inhomogeneities and anisotropies. 
In the present  manuscript  we show that, when considering within the   context of   such  proposals, the 
 characterization  of the spectrum and   bispectrum turn out to be quite different from those  found in the traditional approach, and  in particular,   some of the statistical features,   must be  treated   in a different  way leading  to  some   rather  different   conclusions.
\end{abstract}

%\keywords{Inflation; Quantum  Gravity;  Dynamical Reduction}

%%%%%

\maketitle
\section{Introduction}\label{Intro}

Recent advances in observational cosmology  are allowing a detailed   testing  of our  understanding   of the early universe. In particular, the inflationary paradigm, considered as a one of the most  important  cornerstones of modern  cosmology, is  regarded  as  providing an explanation for  the origin of cosmological structure  \cite{mukhanov1990,hawking1982,starobinsky1982,guth1982,halliwell1985,mukhanov1992}. Indeed,  recent observational data (e.g. WMAP \cite{wmap9}, SDSS \cite{sdss9}, \emph{Planck} \cite{planck}), is in   very good  agreement with the theoretical predictions offered  by  the  inflationary  paradigm. According to  this theory,  the generation of the primordial inhomogeneities,  constituting  ``the seeds of galaxies,'' is also rather image-evoking: The perturbations start as quantum fluctuations of the inflaton field, and  as  the universe   goes  trough an era  of accelerated expansion, the physical wavelength associated with the perturbations is stretched out reaching  cosmological 
scales.

At this point one is  led to treat the quantum fluctuations as classical 
density perturbations; there are several kinds of arguments for such 
consideration. 
However, in most of the original works mentioned above,  the issue  is  only  
dealt implicitly; meanwhile, in more recent works, such as 
\cite{Albrecht, Polarski},  specific  arguments   have  been put forward  based 
on the highly squeezed  nature of  the  field's quantum  state,  or   on 
detailed  decoherence arguments.\footnote{We  must say that  we   do not  find  
those  arguments to be  truly  satisfactory, for reasons  discussed  in detail  
in \cite{sudarsky2009}.} Finally, the traditional picture states that these  
classical perturbations continue evolving into the  cosmic structure 
responsible for galaxy formation, stars, planets and  eventually  life and  
human beings.

However, as  has  been discussed  at length  in  previous works,  the  complete theory must not only  allow  one to  find  expressions  that are in agreement with observations, but also, be able to provide an  explanation of  the   precise  physical   mechanism behind  its predictions. %The inflationary paradigm is an incomplete theory in this sense. 
As originally discussed in \cite{sudarsky2006}, there is a  conceptual difficulty  in the standard explanation for the birth of cosmic structure provided by inflation: The issue    is that   according to   this  account, and  starting  from a highly\footnote{At  the  level of  one part in  $e^{N}$  where $N$  is the number of  e-folds of inflation.} homogeneous and isotropic state that characterizes   the  background  inflaton  and space-time, as well as  the quantum state  of the   perturbations,   the universe  must end in a state with ``actual" inhomogeneities and anisotropies.  Let us  be  a  bit more explicit and  recall  that, when  considering quantum mechanics as a fundamental theory applicable,  in particular,  to the  universe as  a whole,  then one  must regard  any classical description of  the state of any system as a sort of  imprecise characterization of   a complicated quantum mechanical state. The universe that we observe today is clearly well described by an inhomogeneous and anisotropic classical 
state; therefore, such description must be considered as an  imperfect description of an equally inhomogeneous and anisotropic quantum state. Consequently, if we want to consider the  inflationary  account  as  providing the physical  mechanism  for the generation of  the seeds of structure,   such account must contain an explanation for why the quantum state that describes our actual universe does not possess the same symmetries  as  the early quantum state of the universe, which happened to be perfectly symmetric (the symmetry being the homogeneity and isotropy). As  there is nothing in the dynamical evolution (as given by the standard inflationary approach) of the quantum state that can break those symmetries, then we are left with an incomplete theory. In fact, this shortcoming has been recognized by others in the literature \cite{Padmanabhan,Mukhanov,Weinberg,Lyth}.

The detailed discussion of the conceptual problems associated with the inflationary paradigm, and the  proposed   paths  for resolution of the issue  within  schemes  following  standard Quantum Mechanics (e.g., the decoherence program, many-worlds interpretation and the consistent histories approach),  have been discussed  in  detail by some of us and by others in  \cite{susana2013,pearle2007,sudarsky2006,sudarsky2009}. We will not reproduce those arguments here and invite the interested reader to consult those references;  the  above  paragraph  is  meant  only to provide the reader with  a brief indication of the  kinds of issues  that a detailed  analysis  of such questions  involves.\footnote{ In fact,   even if one  wants  to adopt,  say, the many-worlds  interpretation,  the issue  at  hand can be rephrased by  asking  questions  about the precise  nature of the  quantum state that can be taken  as representing our  specific   branch  of the many  worlds. One can also  focus on the issue  of  characterizing in a precise  
mathematical way the  quantities that encode  the  so-called stochastic  aspects, which are often only  vaguely referred  to.  We  will see this   in a very concrete  way  in the  the following.}

The idea that has been presented in previous works \cite{sudarsky2006,sudarsky2006b,sudarsky2007,sudarsky2009,sudarsky2011}, as a possibility to deal with the aforementioned problem, relies on  supplementing the standard inflationary model with  an  hypothesis  involving  the modification of quantum  theory so as to  include  a  spontaneous  dynamical  reduction   of the quantum state  (sometimes referred  as  the self-induced  collapse of the wave function) and  to consider such reduction  as an actual physical process; taking place  independently of observers or measuring devices. 
%We  should note  here the  extensive  amount of   previous work   by  researchers  concerned   with  foundations  of quantum mechanics, about  such ideas  applied to   more  ordinary  situations involving quantum theory\cite{ GRW,  Pearle, Diosi, Penrose}. 
Regarding the situation at hand, the basic scheme is the following: A few $e$-folds after inflation has started, the universe finds itself in an homogeneous and isotropic quantum state, then during the inflationary regime a quantum collapse of the wave function is triggered (by  novel physics  that  could possibly  be  related to quantum gravitational effects),  breaking in the process the unitary evolution of quantum mechanics and also, in   general, the symmetries of the original state.  In our situation the post-collapse state  will   not in general  be  isotropic  nor homogeneous. This  state is  moreover     assumed  to be  sufficiently   peaked in   the relevant   variables  so as to   be   susceptible  to  an approximate  classical  characterization describing  a  universe that includes the inhomogeneities and anisotropies. That  classical description  will  continue   to  evolve  following  the  standard physical processes of post-inflationary cosmology, into  the universe  we have  today.

The idea   of modifying  quantum   mechanics  by incorporating  a  self-induced collapse 
is not a new idea \cite{First Collapse} and there has been a considerably amount of 
research along this lines:  The continuous spontaneous localization (CSL)  model 
\cite{pearle1989}, representing  a   continuous  version of the Ghirardi-Rimini-Weber 
(GRW) model \cite{ghirardi1985}, and the ideas of Penrose \cite{penrose} and Di\'osi 
\cite{diosi} regarding gravity as the main agent responsible for the collapse, are among 
the main programs proposed to describe the physical mechanism of a self-induced 
collapse of the wave function. For more recent examples  see Refs. 
\cite{weinberg2011,bassi2003}.
In fact, the implications of applying the CSL model to the inflationary scenario have been 
studied in Refs. \cite{pedro,jmartin} and \cite{tpsingh} leading to interesting results   
constraining  the parameters of the model in terms of   the parameters  characterizing  
the inflationary  model.

On the other hand,  the statistical analyses  in  Refs. \cite{adolfo,susana2012} show  
how the predictions of the simple collapse schemes,  used  in  previous  works,  can be  
confronted with recent data from the Cosmic Microwave Background (CMB), including 
the 7-yr release of WMAP \cite{wmap7} and the matter power spectrum measured 
using LRGs by the Sloan Digital Sky Survey \cite{sdssb}. In fact, results from those 
analyses indicate that  while  several schemes or ``recipes'' for the collapse are 
compatible with the observational data,   others  are not, allowing to establish 
constraints on the free parameters of the schemes. Those  works serve to underscore 
the point   that,   in addressing the conceptual issues confronting  the inflationary 
paradigm, one  is not   only dealing   with ``philosophical issues,'' but that   these  have  
impact in the theoretical predictions. While,   at the same time,  the conclusions  drawn  
can  lead  to  important insights,   as well as  a better  understanding  of the nature  of  
those  predictions  and novel ways to  consider the relation  between  such   
predictions and  the observations.

In this  work,  we will be primarily concerned with the characteristics of   the ``primordial bispectrum''   although   we  will  also make  some   interesting 
observations  regarding the   primordial  spectrum itself.  In the inflationary paradigm, 
the primordial bispectrum has been regarded as  one of the indicators  characterizing   
possible  primordial ``non-Gaussianities.''  By non-Gaussianity, one    generically refers to any  
deviations in the observed fluctuations from the random field of linear, Gaussian, 
curvature perturbations. 
It is commonly believed that the study of non-Gaussianities will play a leading role in 
furthering our understanding of the physics of the very early universe that created the 
primordial seeds for large-scale structures \cite{komatsu2009}. As a matter of fact, the 
shape and amplitude of the bispectrum is  currently being  used  in order  to discriminate 
among various inflationary models. The amplitude of non-Gaussianities is 
characterized in terms of a dimensionless  parameter $f_{\textrm{NL}}$ (defined in 
Sec. \ref{diferencias}). Distinct models of inflation predict different values for  
$f_{\textrm{NL}}$. The amount of  non-Gaussianity from simple inflationary models, 
which are based on a single  scalar filed  with slow-roll  inflationary regime, is expected  
to be very small \cite{salopek1990,salopek1991,falk1992,gangui1993,acquaviva2002,maldacena2002}; 
however, a very large class of more general models, such as   those  with  multiple scalar fields,   special features in  the inflaton potential, non-canonical kinetic terms, deviations from Bunch-Davies vacuum, and  others (see Ref. \cite{bartolo2004} and references therein for a review ),  are thought to be able  to generate substantially higher levels  of non-Gaussianity.\footnote{For more details and derivations regarding non-Gaussianity, we refer the reader to the comprehensive review by Komatsu \cite{komatsu2003}, Bartolo et al. \cite{bartolo2004} and others \cite{Yadav,Liguori,Komatsu2001}.}

In this article, we will consider the simplest slow-roll inflation model, within  the  approach   involving  the hypothesis of ``the collapse of the wave function.''   We will  focus on the  simplest  collapse  scheme  involving a  single  instantaneous  collapse per 
mode  proposal,  although  several of the results   will  be   valid  in more  general  
schemes.   We   will  also  use  some  previous results \cite{alberto,sigma}, which 
indicate  that certain  intrinsic  non-linearities  associated  with the collapse of the wave 
function in a gravitational context,  generically induce correlations between the random 
variables characterizing the collapse of the wave function in the different  modes. In 
other words, the collapse of the wave function, cannot operate in an exactly mode 
independent manner, and  that whenever the collapse takes place, it  leads  to 
correlations   in the state  of  the  different modes \cite{alberto}. In fact,  the  basic  
aspects of  the  collapse-induced correlation between the modes and its particular 
signature in the CMB angular power spectrum has been analyzed in Ref. \cite{sigma}. In 
the present letter, we will be primarily interested in analyze the imprint in the CMB 
primordial bispectrum, by taking into account the correlation between the modes 
generated by the self-induced collapse. This  analysis is  made possible because, in our 
approach,  we must use  explicit   variables  characterizing  the random aspects  that   
determine   the primordial bispectrum. 

Therefore, although  we agree   with the  prevailing view,  that the  primordial 
bispectrum offers us a  valuable  observational tool to examine  some physical 
processes  that  were relevant in the early universe, the statistical aspects  that are 
involved   in the  comparison of theory and observations,  must, in  our view,   be 
addressed   in  a different manner.

At this point,   and in order to avoid any  misconceptions,   we  must  warn the reader  familiar with the subject  that we will not   be   discussing  here   anything related  to the quantum  three-point function, but  that we  are  focusing on  the quantity that  the  observers  actually measure, and   which, in   our view,  differs  in  essential  ways  from the former. In particular, we will assume a highly Gaussian primordial perturbation, and after including the self-induced collapse of the inflaton's wave function, we will obtain an expression for the observed bispectrum that will be derived from the 6th. momentum of  distribution associated to the coefficients $a_{lm}$ (corresponding to the harmonic decomposition of the temperature anisotropies on the CMB).  We note that, in the standard approach, the assumption of an initial Gaussian perturbation would lead to a zero value for the bispectrum, up to its cosmic variance. Therefore, for all practical purposes, followers of the traditional inflationary paradigm may reinterpret our result as a new estimate for the bispectrum variance that is different in shape from  the traditional one. In particular,  some of the differences arises as  a result of  having included the collapse of the wave function. However from the conceptual point of view, our approach and the standard one, differ significantly, hence, what is identified at the theoretical  level  as  the bispectrum within our approach  is not the same as  what is   called  by that  name  in  the traditional approach. The differences are subtle but  very  important, and we encourage the reader to consult Ref. \cite{susana2013}, in order to avoid  the  misunderstandings  that  result from   considering  our  statistical  analysis   without appropriately  considering   the  difference  in the whole approach.

%  In addition to the primordial bispectrum, we will analyze a new quantity (previously introduced in Ref. \cite{susana2013}) that offers two advantages with respect of the former. 
% First, the observational and theoretical notions are clearly separated; this contrasts with the common treatment for obtaining an observational value for the amplitude and shape of the bispectrum. Second, it serves to illustrate that maintaining clear distinctions between each of the averages involved (quantum averages, ensemble averages, time-space averages and orientation averages)  yields a  different  set of predictions for the observational quantities.
% As  we  will  find,  in the traditional picture the   expected value for the new quantity vanishes exactly, while working in the approach  based  on  collapse framework, one is lead  to  expect a  non-vanishing value  for this  quantity. 

 The  reminder of the  paper is organized as follows: In Section \ref{review}, we start by reviewing the ideas and technical aspects of our proposal, in particular we focus on how to implement  the collapse of the wave function during the inflationary universe. Then, in Section \ref{seccolapsobispec}, we derive an analytic estimate of the  expected value  of the  observed  primordial bispectrum within our approach. Afterwards, in Section \ref{diferencias}, we discuss the main differences between the collapse proposal and the standard approach regarding  the estimates  of  the  primordial bispectrum and its statiscal aspects.  Finally, in Section \ref{concl}, we end with a discussion of our conclusions. 

 %Section \ref{nuevacantidad} contains a detailed analysis for a novel observational quantity that allows us to 
% differentiate between the two approaches, both, at the quantitative and qualitative level. In Section \ref{comparisons}, we present a discussion regarding the comparisons between our theoretical prediction and the observational data. 
%We have also included 
%two Appendixes \ref{calculos}, \ref{apendiceb} that present   the details of the  more lengthy calculations appearing  in Secs. \ref{seccolapsobispec} and \ref{nuevacantidad}. 
%respecitvelly.

Regarding conventions and notation, we will be using a $(-,+,+,+)$ signature for the space-time 
metric. The prime over the functions  $f'$ denotes derivatives with respect to the 
conformal time $\eta$. We will use units where $c=\hbar=1$ but will keep the gravitational constant 
$G$.

\section{Brief review of the collapse proposal within the inflationary universe}\label{review}

In this section we will present a brief review of the collapse proposal which has been exposed in great 
detail in previous works \cite{sudarsky2006,alberto,sigma,adolfo,adolfo2010,alberto2012}. The main 
purpose is to present  the central ideas behind the proposal in order to make the presentation as self-
contained as possible. 
%Here, we will expose the manner in which the collapse 
%of the wave function enters into the picture of the inflationary universe. 
First, we will still describe the formalism in a concise manner;  although     
we  will  not  be  applying  all   its details  due to  its intrinsic  complexity,   it  serves  to   explain  
the  general idea  and   show explicitly how the collapse generates  correlations between different modes. On the 
other hand,  we  believe that  the   analysis  we  will preform,  would have analogous   counterparts  in 
other  much  more developed  approaches involving dynamical collapse  theories,  such  as the    GRW \cite{ghirardi1985}  or CSL \cite{pearle1989}  proposals,    and   even  in approaches that rely on 
applying Bohmian Mechanics  to the cosmological   problem  \cite{valentini,pintoneto}.  For examples 
using CSL  in this context  see \cite{jmartin, tpsingh, pedro}.

\subsection{Semiclassical self-consistent configuration}\label{SSC}

Before proceeding with the technical aspects, we wish to discuss  the way the gravitational sector and  the matter fields are   treated in our  approach. As we have  not yet at our disposal  a fully workable  and satisfactory theory of quantum gravity, we will rely on the ``semiclassical gravity'' approach, which  naturally   we  take  only as   an effective  setting  rather than  something that can  be  considered as  a  fundamental theory. The inflationary period is assumed to start at energy scales smaller than the Planck mass ($\sim 10^{-2}-10^{-3} M_P$), thus,  one  can  expect  that the semiclassical approach is a suitable approximation  for something that, in principle,  ought to  be treated   in   a precise  fashion  within a quantum theory  of  gravity. The semiclassical framework is characterized by Einstein semiclassical equations $G_{ab} = 8 \pi G \bra \hat{T}_{ab} \ket$, which  allow to relate the quantum   treatment of  the degrees of freedom  associated   with  the matter fields to the classical description of gravity in terms of the metric.\footnote{As  wee sill see during the collapse, the  semiclassical approximation will not remain 100\% valid, because the quantum collapse or jump of the  quantum state  we   will have $\nabla_a \bra \hat{T}^{ab} \ket \neq 0$, while $\nabla_a G^{ab} = 0$. However, as we will be only interested in the states \emph{before} and \emph{after} the collapse, this  breakdown of the semiclassical approximation would not be important for our present work.} The use of such semiclassical picture has two main conceptual advantages:

First, the description and treatment of the metric is always ``classical." As a consequence there is no issue with the ``quantum-to-classical transition" in the characterization   of space-time.  Thus  we     will not   need to justify going from ``metric operators" (e.g. $\hat \Psi$) to classical metric variables (such as $\Psi$).  The fact that the space-time remains classical is particularly important in  the context  of models  involving dynamical reduction of the wave function,  as   such    ``collapse or reduction" is  regarded   as {\it a  physical process taking place in time}  and,  therefore,   it is  clear  that  a  setting allowing  consideration of  full space-time  notions  is  preferred  over, say,  the  ``timeless"  settings usually  encountered  in canonical  approaches to quantum gravity (for  some basic references on ``the problem of time  on quantum gravity''  see Ref. \cite{TIME}).

Second, it allows to present a transparent picture of how the inhomogeneities and anisotropies are born from the quantum collapse: the initial state of the universe (i.e. the one characterized by a few $e$-folds after inflation has started) is described by the homogeneous and isotropic Bunch-Davies vacuum, and the equally homogeneous and isotropic classical Friedmann-Robertson-Walker space-time. Then, at a later stage, the quantum state of the matter fields reaches a regime   whereby  (following the general ideas advocated  by  R .Penrose) the corresponding state for the gravitational degrees of freedom are forbidden, and a quantum collapse of the matter field wave function is triggered by some unknown physical mechanism. In this manner, the state resulting from the collapse needs not to share the same symmetries as the initial state. After the collapse, the gravitational degrees of freedom are assumed to be, once more, accurately described by Einstein semiclassical equation. However, as $\bra \hat{T}_{ab} \ket$ for the new state needs not to have the symmetries of the pre-collapse state, we are led to a geometry that generically will no longer be homogeneous and isotropic.

Another advantage of the semiclassical framework is that the universe can be described by what we have named ``the semiclassical self-consistent configuration'' (SSC), which was originally presented in Ref. \cite{alberto}. In the following, we provide a brief description of such proposal.

The SSC considers  a space-time  geometry characterized by a  classical  space-time  metric  and  a standard quantum field theory  constructed  on that  fixed space-time background, together with a particular quantum  state in that construction,  such that the  semiclassical   Einstein's  equations hold.   More precisely, we  will  say that the set 

\beq
\left\lbrace g_{\mu\nu}(x),\hat{\varphi}(x),\hat{\pi}(x),\mathscr{H},\vert\xi\rangle\in\mathscr{H}\right\rbrace
\eeq
corresponds to  a SSC if and only if $\hat{\varphi}(x)$, $\hat{\pi}(x)$ and $\mathscr{H}$  correspond to a quantum field theory constructed over a space-time  with  metric  $g_{\mu\nu} (x)$, and  the  state $\vert\xi\rangle$ in the Hilbert space $\mathscr{H}$  is  such that

\beq\label{Mset-up}
G_{\mu\nu}[g(x)]=8\pi G\langle\xi\vert \hat{T}_{\mu\nu}[g(x),\hat{\varphi}(x),\hat{\pi}(x)]\vert\xi\rangle,
\eeq
for all the points in the space-time manifold.

The former  description  is  thought to  be  appropriate  in the   regime of interests  except  at those  specific   times  when  a  collapse   takes place (because of the reasons mentioned in footnote 4). 
 If  we   view   the   SSC  as the   reasonable  characterization of  situations   suitable to be treated  within the  semi-classical    approximation,   and consider  the   various   post-Planckian   cosmological  epochs   as    susceptible to such   description,  we must  consider  collapses as   taking the universe  from  one  such  regime   say  SSC-i  to another say  SSC-ii.
 
The relation between the SSC and the collapse process can be described heuristically  in the following  way: first, within the Hilbert space associated to the  given SSC-i, one can consider that a sudden 
jump $\vert\xi^{\textrm{(i)}}\rangle\to\vert\zeta^{\textrm{(i)}}\rangle_{\textrm{target}}$ ``is 
about to occur,'' with both $\vert\xi^{\textrm{(i)}}\rangle$ and $\vert\zeta^{\textrm{(i)}}
\rangle_{\textrm{target}}$ in $\mathscr{H}^{\textrm{(i)}}$.  However, in general, the 
set $\{g^{\textrm{(i)}},\hat{\varphi}^{\textrm{(i)}},\hat{\pi}^{\textrm{(i)}}, \mathscr{H}^{\textrm{(i)}},\vert\zeta^{\textrm{(i)}}\rangle_{\textrm{target}}\}$ will not represent  a new SSC. In order to 
describe a reasonable picture,  as presented in Ref.  \cite{alberto},  one needs to relate  the  state $
\vert\zeta^{\textrm{(i)}}\rangle_{\textrm{target}}$ with another one $\vert\zeta^{\textrm{(ii)}}
\rangle$ ``living'' in a new Hilbert space $\mathscr{H}^{\textrm{(ii)}}$ for which
 $\{g^{\textrm{(ii)}},\hat{\varphi}^{\textrm{(ii)}},\hat{\pi}^{\textrm{(ii)}}, \mathscr{H}^{\textrm{(ii)}},\vert\zeta^{\textrm{(ii)}}\rangle\}$ is a valid SSC;  the   one  denoted by SSC-ii. 
Consequently, one needs to determine first the ``target'' (non-physical) state in $\mathscr{H}^{\textrm{(i)}}$ to which the initial state is ``tempted'' to jump, sort of speak, and after that, one 
can relate such target state with a corresponding  state  in the Hilbert space of the SSC-ii. One then 
considers that the target state is chosen stochastically, guided by the quantum uncertainties of 
designated  field  operators, evaluated on the initial state $\vert\xi^{\textrm{(i)}}\rangle$ and  at 
the time of collapse.

In Ref. \cite{alberto} a prescription for identifying two different SSC's, during the collapse has been 
introduced; the idea is the following: Assume that the collapse takes place along a Cauchy hypersurface $\Sigma$. A transition from the physical state $\vert\xi^{\textrm{(i)}}\rangle$ in $
\mathscr{H}^{\textrm{(i)}}$ to the physical state $\vert\zeta^{\textrm{(ii)}}\rangle$ in $
\mathscr{H}^{\textrm{(ii)}}$ (associated to a certain  target \textit{non-physical} state $\vert
\zeta^{\textrm{(i)}}\rangle_{\textrm{target}}$ in $\mathscr{H}^{\textrm{(i)}}$) will occur  in a way 
that

 \beq\label{recipe.collapses}
_\textrm{target}\langle\zeta^{\textrm{(i)}}\vert \hat{T}^{\textrm{(i)}}_{\mu\nu}[g^{\textrm{(i)}}, \hat{\varphi}^{\textrm{(i)}},\hat{\pi}^{\textrm{(i)}}]\vert\zeta^{\textrm{(i)}}\rangle_{\textrm{target}} \big|_{\Sigma}= \langle\zeta^{\textrm{(ii)}}\vert \hat{T}^{\textrm{(ii)}}_{\mu\nu}[g^{\textrm{(ii)}}, \hat{\varphi}^{\textrm{(ii)}},\hat{\pi}^{\textrm{(ii)}}]\vert\zeta^{\textrm{(ii)}}\rangle \big|_{\Sigma}, 
\eeq
this is, in such a  way  that  the expectation value of the energy momentum tensor, associated to the states $\vert\zeta^{\textrm{(i)}}\rangle_{\textrm{target}}$ and $\vert\zeta^{\textrm{(ii)}}\rangle$ evaluated on the Cauchy hypersurface $\Sigma$, coincides. Note that the left hand side  in the expression  above is meant  to be constructed from the elements of the SSC-i (although $\vert\zeta^{\textrm{(i)}}\rangle_{\textrm{target}}$ is not really {\it the state} of the SSC-i), while the right hand side correspond to quantities  evaluated  using the SSC-ii.

For the case analyzed in \cite{alberto},  the SSC-i  corresponded to the homogeneous and  isotropic  space-time, i.e. $\Psi_{\nk}=0$ and the state of the quantum field corresponding to the Bunch-Davies vacuum, while the  SSC-ii  corresponded  to an excitation of  a single mode $\vec{k_0}$, characterized  by $\Psi= F(\eta) \cos (\vec{k_0} \cdot \vec x )$, with a particular quantum state for the  inflaton  field. The energy momentum tensor associated to that  state is compatible with this  space-time  metric according to  the SSC recipe. As  shown   in \cite{alberto}, the main point is that,  when the  SSC-ii  corresponds to  a curvature perturbation $\Psi$  including  spatial  dependences   with  wavenumber $\vec{k_0}$, the normal  modes  of the   corresponding Hilbert space    $\mathscr{H}^{\textrm{(ii)}}$,  which  would  otherwise  be characterized  by the typical   spatial dependence $e^{i \vec{k}.\vec x}$,  and  would  be called  the $\vec{k}$  modes,   would  now  contain also  corrective  contributions   of the form $e^{i( \vec{k}\pm \vec{k_0}).\vec x}$ simply because the space-time   is  not  exactly  homogeneous and isotropic.

Conversely, this implies   that  if  the  energy momentum tensor does  not lead to the excitation  of  other  modes  besides the $ \pm \vec{k_0}$ in the Newtonian potential $\Psi$  (as  required   for equation \ref{Mset-up} to hold),  then,  the  state of the quantum field  needs to  be excited  not just  in the modes $\pm \vec{k_0}$,  but also  in the  modes $ \pm  n \vec{k_0}$,  with $n$ integer (with the degree  of excitation  decreasing with increasing $n$). In other words, the post-collapse state is characterized by

\beq
|\zeta^{\textrm{(ii)}} \ket = \ldots  |\zeta^{\textrm{(ii)}}_{-2\nk_0}  \ket \otimes |\zeta^{\textrm{(ii)}}_{-\nk_0} \ket \otimes |\zeta^{\textrm{(ii)}}_{\nk_0} \ket \otimes |\zeta^{\textrm{(ii)}}_{2\nk_0} \ket  \ldots
\eeq

This  situation    can be   understood  as  arising  from the intrinsic  non-linearity  of the SSC construction,  which  corresponds   in a sense   to   a  relativistic  version of the  Newton-Schr\"odinger  system \cite{Newt Sch}. 

The details  of self-consistent formalism  can be consulted in Ref.  \cite{alberto}. We  will not use such full fledged formal treatment here, in  part  because the analysis  becomes extremely cumbersome even in the treatment of a single mode of the inflationary field. Thus, 
%even though, it is in principle 
%possible to use such detailed formalism, 
when studying the CMB bispectrum, the task  would  quickly become a practical impossibility. Instead, 
we will focus on the   simpler  pragmatical approach first proposed in \cite{sudarsky2006}  but keep in  
mind  the lessons  learned  from the SSC formalism. In the next subsection, we will describe how to 
implement the  basic results of the SSC proposal in a practical way.

% 
% In those circumstances,  one can not longer  argue that  at the  time  when   these  higher  modes collapse, there  should be  complete  symmetry in the {\it a priori}  statistical  distribution of the post-collapse   parameters  of the state.
% 
% That is,   there seems to be ample  justification to assume that  the  value of the  random   parameter $x_{n \vec{k_0}}$ ($n\not = 1$), associated with the expectation value in the post-collapse state of the field,  would not  be  completely independent from the   value taken   by the random parameters  $x_{\vec{k_0},y}$. Focussing for simplicity on the  modes,   which in view of the  preceding   discussion one  expects to  be  more closely correlated, we  limit  hereafter consideration to the  case  $n=2$, and  characterize  the level of correlation by  an unknown  quantity  we label $ \varepsilon$. We  will show in detail how to implement these ideas  in  Section \ref{bla}.
% 

\subsection{The primordial curvature perturbation and the self-induced collapse}

We proceed now to introduce the details of the  simplest  collapse proposal. The starting point will be  the same as the one underlying the standard slow-roll inflationary model; this is one considers  the action of a scalar field (the inflaton) minimally coupled to gravity, 

\beq\label{accioncolapso}
S[\phi,g_{ab}] = \int d^4x \sqrt{-g} \bigg( \frac{1}{16 \pi G} R[g] - \frac{1}{2} \nabla_a \phi \nabla_b \phi g^{ab} - V[\phi] \bigg).
\eeq

This  leads to  the   Einstein's field equations $G_{ab} = 8 \pi G T_{ab}$  with  $T^a_b$ given by:

\beq
T^a_b = g^{ac} \partial_c \phi \partial_b \phi + \delta^a_b \left( -\frac{1}{2} g^{cd}\partial_c \phi \partial_d \phi - V[\phi] \right).
\eeq

and the  field  equations  for the  scalar field:
\beq
  \nabla^a  \nabla_a \phi + \frac {\partial  V[\phi]}
   {\partial\phi }
   =0 
\eeq

The next step is to split the metric and the scalar field into a background plus perturbations $g_{ab} = g_{ab}^{(0)} + \delta g_{ab}$, $\phi = \phi_0 + \dphi$. The background is represented by a spatially flat FRW space-time with line element $ds^2 = a(\eta)[-d\eta^2 + \delta_{ij}dx^i dx^j]$ and the homogeneous part of the scalar field $\phi_0 (\eta)$. 

The scale factor corresponding to the inflationary era is $a(\eta) \simeq -1/(H \eta)$ with $H$ the Hubble factor defined as $H \equiv \partial_t a/a$, thus $H \simeq$ const. During inflation $H$ is related to the inflaton potential as $H^2 \simeq (8 \pi G /3) V$. The scalar field $\phi_0(\eta)$ is in the slow-roll regime, which means that $\phi_0' \simeq -(a^3/3a') \partial_\phi V$. The slow-roll parameter defined by $\epsilon \equiv \frac{1}{2} M_P^2 (\partial_\phi V/V)^2$ is considered to be $\epsilon \ll 1$;  $M_P$ is the reduced \emph{Planck} mass defined as $M_P^2 \equiv 1/(8 \pi G)$.
We will set $a=1$ at the present cosmological time; while we assume that the inflationary period ends at a conformal time $\eta_\star \simeq -10^{-22}$ Mpc.

%%%%%%%
%%%%%%

Next,  we focus on the perturbations. Ignoring the vector and tensorial perturbations, and working  in the conformal Newtonian gauge, the perturbed space-time is represented by

\beq
ds^2 = a(\eta)^2 [-(1+2\Phi) d\eta^2 + (1- 2\Psi)\delta_{ij}dx^idx^j],
\eeq
with $\Phi$ and $\Psi$ functions of the space-time coordinates $\eta,x^i$. If we assume that there is no anisotropic stress, Einstein's equations to first order in the perturbations lead to $\Phi = \Psi$; thus, combining Einstein's equations yield: 

\beq\label{master}
\nabla^2 \Psi = 4\pi G \phi_0' \dphi' =- \sqrt{\frac{\epsilon}{2}} \frac{aH}{ M_P} \dphi',
\eeq
where in the second equality we used Friedmann's equations and the definition of the slow-roll parameter.

Next, we consider the quantization of the theory. As  mentioned  above,  we will  work  within the  collapse-modified  semiclassical gravity 
setting. In particular, we will quantize the fluctuation of the inflaton field $\dphi (\x,\eta)$,  but not the  metric perturbations. For simplicity, we will work with the rescaled field variable $y=a\dphi$. One then proceeds to expand the action \eqref{accioncolapso} up to second order in the rescaled variable (i.e. up to second order in the scalar field fluctuations)

\beq\label{acciony}
\delta S^{(2)}= \int d^4x \delta \mathcal{L}^{(2)} = \int d^4x \frac{1}{2} \left[ y'^2 - (\nabla y)^2 + \left(\frac{a'}{a} \right)^2 y^2 - 2 \left(\frac{a'}{a} \right) y y' \right].
\eeq
The canonical momentum conjugated to $y$ is $\pi \equiv \partial \delta \mathcal{L}^{(2)}/\partial y' = y'-(a'/a)y=a\dphi'$.

In order to avoid distracting infrared divergences, we set the problem in a finite box of side $L$. At the end of the calculations we can take the continuum limit by taking $L \to \infty$. The field and momentum operators are decomposed in plane waves

\beq
\hat{y}(\eta,\x) = \frac{1}{L^3} \sum_{\nk} \hat{y}_{\nk} (\eta) e^{i \nk \cdot \x}  \qquad \hat{\pi}(\eta,\x) = \frac{1}{L^3} \sum_{\nk} \hat{\pi}_{\nk} (\eta) e^{i \nk \cdot \x},
\eeq
where the sum is over the wave vectors $\vec k$ satisfying $k_i L=
2\pi n_i$ for $i=1,2,3$ with $n_i$ integer and $\hat y_{\nk} (\eta) \equiv y_k(\eta) \ann_{\nk} + y_k^*(\eta)
\cre_{-\nk}$ and  $\hat \pi_{\nk} (\eta) \equiv g_k(\eta) \ann_{\nk} + g_{k}^*(\eta)
\cre_{-\nk}$. The function $y_k(\eta)$ satisfies the equation:

\beq\label{ykmov}
y''_k(\eta) + \left(k^2 - \frac{a''}{a} \right) y_k(\eta)=0.
\eeq
To complete the quantization, we have to specify the mode  solutions of \eqref{ykmov}.  The canonical commutation relations between $\hat y$ and $\hat \pi$, will  give $[\hat{a}_{\nk},\hat{a}^\dag_{\nk'}] = L^3 \delta_{\nk,\nk'}$, when $y_k(\eta)$ is  chosen  to satisfy $y_k g_k^* - y_k^* g_k = i$ for all $k$ at some time $\eta$.

The remainder of  the  choice of $y_k(\eta)$ corresponds to the so-called Bunch-Davies  (BD) vacuum, which is characterized by

\beq
y_k(\eta) = \frac{1}{\sqrt{2k}} \left( 1- \frac{i}{\eta k} \right) e^{-ik\eta}, \qquad g_k(\eta) = -i \sqrt{\frac{k}{2}} e^{-ik\eta}.
\eeq
There is certainly  some   arbitrariness  in  selection of a  natural  vacuum state,     but it seems  clear that any such  natural choice would be spatially a homogeneous and isotropic state. The  BD vacuum  certainly is a homogeneous and isotropic state as can be seen by evaluating directly the action of a translation or rotation operator on the state.

From $G_{ab} = 8 \pi G \bra \hat{T}_{ab} \ket$ and \eqref{master} it follows  that

\beq\label{master2}
\Psi_{\nk} (\eta) = \sqrt{\frac{\epsilon}{2}} \frac{H}{M_P k^2} \bra \hat{\pi}_{\nk} (\eta) \ket.
\eeq
It is clear from Eq. \eqref{master2} that if the state of the field is the vacuum state, the metric perturbations vanish, and, thus the space-time is homogeneous and isotropic.

The self-induced collapse model is based on considering that the collapse operates very similar to a  kind of self-induced ``measurement" (evidently, there is no external observer or detector involved). In considering  the operators  used to characterize the post-collapse  states,  it seems  natural therefore  
to focus  on Hermitian operators,  which in ordinary quantum mechanics are the ones susceptible of direct measurement. 
We thus  separate $\hat y_{\nk} (\eta)$ and $\hat \pi_{\nk} (\eta)$ 
into  their  ``real and imaginary parts" $\hat y_{\nk} (\eta)=\hat y_{\nk}{}^R (\eta) +i \hat y_{\nk}{}^I (\eta)$ and $\hat \pi_{\nk} (\eta) =\hat \pi_{\nk}{}^R (\eta) +i \hat \pi_{\nk}{}^I (\eta)$ . The point is that the operators $\hat y_{\nk}^{R, I} (\eta)$ and $\hat \pi_{\nk}^{R, I} (\eta)$ are  hermitian. Thus,\footnote{$\mathcal{R}[z]$ 
denotes the real part of 
$z \in \mathbb{C}$} $\hat{y}_{\nk}^{R,I} (\eta) = \sqrt{2} \mathcal{R}[y_k(\eta) \hat{a}_{\nk}^{R,I}]$, $\hat{\pi}_{\nk}^{R,I} (\eta) = \sqrt{2} \mathcal{R}[g_k(\eta) \hat{a}_{\nk}^{R,I}]$,
where $\hat{a}_{\nk}^R \equiv (\hat{a}_{\nk} + \hat{a}_{-\nk})/\sqrt{2}$, $\hat{a}_{\nk}^I \equiv -i (\hat{a}_{\nk} - \hat{a}_{-\nk})/\sqrt{2}$.
 The commutation relations for the $\hat{a}_{\nk}^{R,I}$ are non-standard

\beq\label{creanRI}
[\hat{a}_{\nk}^R,\hat{a}_{\nk'}^{R \dag}] = L^3 (\delta_{\nk,\nk'} + \delta_{\nk,-\nk'}), \quad [\hat{a}_{\nk}^I,\hat{a}_{\nk'}^{I \dag}] = L^3 (\delta_{\nk,\nk'} - \delta_{\nk,-\nk'}),
\eeq
with all other commutators vanishing.

Following the a  line of   thought  described  above, we  assume that the collapse is somehow analogous to an imprecise measurement\footnote{An imprecise measurement of an observable is 
one in which one does not end with an exact eigenstate  of that observable, but  rather with a state that is  only peaked around the eigenvalue. Thus, we could consider measuring a  
certain particle's position and momentum so as to end up with a state that is a wave packet with both position and momentum defined to a limited extent and, which certainly, does not  entail a conflict with Heisenberg's uncertainty bound.} of the operators $\hat y_{\nk}^{R, I}(\eta)$ and $\hat \pi_{\nk}^{R, I}(\eta)$. The rules according to which the collapse  is assumed to  happen are guided by simplicity and naturalness.

In particular,   as  we  are  taking the view that a collapse  effect on  a state is analogous  to some  sort of  approximate measurement, we  will postulate   that  after the  collapse, the expectation values of 
the field and momentum operators  in each mode  will  be related to the uncertainties  of the  initial  state. For the purpose of this  work  we will  work with a particular collapse scheme called  the  
\emph{Newtonian} collapse scheme which is given by\footnote{In previous works, we have analyzed other collapse schemes such as the \emph{independent} scheme and the 
\emph{Wigner} scheme. See Refs. \cite{sudarsky2006,adolfo,susana2012} for detailed analyses of the collapse schemes.}

\begin{equation}
\langle {\hat{y}_{\nk}^{R,I}
(\eta^c_k)} \rangle_\Theta = 0 , 
%x^{R,I}_{\nk,1} \sqrt{\fluc{\hat{y}^{R,I}_{\nk}}_0} = \lambda_1 x^{R,I}_{\nk,1}|y_k(\eta^c_k)|\sqrt{ L^3/2},
\label{momentito}
\end{equation}

\begin{equation}
\langle {\hat{\pi}_{\nk}{}^{R,I}
(\eta^c_k)}\rangle_\Theta
= x^{R,I}_{\nk}\sqrt{\fluc{\pyRI_{\nk}}
_0} ,
=  x^{R,I}_{\nk}|g_k(\eta^c_k)|\sqrt{L^3/2},
\label{momentito1}
\end{equation}
where $\tc$ represents the \emph{time of collapse} for each mode. In the vacuum state,  $\hat{\pi}_{\nk}$ is distributed according to a Gaussian wave function centered at $0$ with spread  
$\fluc{\hat{\pi}_{\nk}}_0$.  The motivation for choosing such scheme is two-folded. First, the calculations performed for this scheme are relatively easier to handle than  other schemes  that have been  considered, and second, in Eq. 
\eqref{master2} the variable that is directly related with the Newtonian Potential $\Psi$ is the expectation value of $\hat \pi$; therefore, it seems natural to consider that the variable 
affected at the time of collapse is $\bra  \hat \pi_{\nk} (\tc) \ket$  while 
there is  no change in the   mean  value  of the  conjugate variable,  thus  
$\bra \hat{y}_{\nk} (\tc) \ket = 0$.

The  random variables $x_{\nk}^{R,I}$   represent  values  selected randomly  from  a  distribution  that in principle might possess ``small non-Gaussian'' features, in the sense that, as discussed in Sec. \ref{SSC}, the collapse induces correlations between the modes. In other words, there seems to be a clear justification to presume that the value of the random parameter $x_{n\nk}^{R,I}$ (with $n \neq 1$) would not be completely independent from the value taken by the random parameters $x_{\nk}^{R,I}$.  Heuristically, the point is that  once  the   mode $\vec k$  has collapsed,  the modes corresponding to the higher harmonics
$n\vec k$ become excited as  explained above (with the  largest  effect   for  $ n=2$),  and  just as it  occurs  in  the  situations  involving  the ``stimulated  emission of  photons'' (such as  the  one  underlying  the functioning of  lasers),  a  mode  that is  already  excited  in certain way,   has a higher  propensity to  becoming   excited in that manner  than it  would be  otherwise.  
Thus,  focusing for simplicity on the modes, which in view of the preceding discussion one expects to be more closely correlated, we limit hereafter consideration to the case $n = 2$, and characterize the level of correlation by an unknown quantity we label $\lambda < 1$. Specifically, we will assume that  the average over possible outcomes of the product of two random variables $x_{\nk}^{R,I}$ are characterized by

\begin{subequations}\label{equisri}
\beq\label{equisr}
\overline{x_{\nk}^R x_{\nk'}^R} = \delta_{\nk,\nk'} + \delta_{\nk,-\nk'} + \lambda ( \delta_{\nk,2 \nk'} + \delta_{\nk,-2 \nk'} + \delta_{2 \nk,\nk'} +\delta_{-2\nk,\nk'} ),
\eeq
\beq\label{equisi}
\overline{x_{\nk}^I x_{\nk'}^I} = \delta_{\nk,\nk'} - \delta_{\nk,-\nk'} + \lambda ( \delta_{\nk,2 \nk'} - \delta_{\nk,-2 \nk'} + \delta_{2 \nk,\nk'} -\delta_{-2\nk,\nk'} ).
\eeq
\end{subequations}

On the other hand, we must emphasize that our universe  corresponds to a \textbf{single realization} of these random variables, and thus  each of these quantities $ x^{R}_{\nk}$, $ x^{I}_{\nk}$ has a  single specific value. It is  clear that  even though  we  will not do that here,  one could   also investigate  how  the statistics  of   $x_{\nk}^{R},x_{\nk}^{I}$ might be affected in detail  by the physical  process of the collapse. The statistical  aspects  characterizing   these  quantities can  be studied   using as  a tool an imaginary ensemble of ``possible universes,'' but  as  explained in \cite{susana2013} we  should,   in principle, distinguish   those   from  the statistical  characterization of  such  quantities for the particular  universe  we inhabit; we will discuss these and other aspects in  more detail  the next sections.

The next step is to find an expression for the evolution of the expectation values of the field operators at  all  times. 
%The  quantity
%$d^{R,I}
%_{\nk} \equiv \langle\Theta|\ann_{\nk}^{R,I}
%|\Theta\rangle $,   determines   the  expectation value of the field and momentum  operator for the mode  $\nk$ at all times after the collapse. That is
%\beq\label{expeceta}
%\bra \hat{y}_{\nk}^{R,I} (\eta) \ket_{\Theta} = \sqrt{2} \mathcal{R}[y_k(\eta) d_{\nk}^{R,I}], \qquad \bra \hat{\pi}_{\nk}^{R,I} (\eta) \ket_{\Theta} = \sqrt{2} \mathcal{R}[g_k(\eta) d_{\nk}^{R,I}],
%\eeq
%correspond to expectation values at any time after the collapse in the post-collapse state $| \Theta \ket$. One can then relate the value of $d_{\nk}^{R,I}$ with the value of the expectation 
%value of the field operators at the time of collapse (i.e. exactly after the instantaneous  collapse) $\bra \hat{y}_{\nk}^{R,I} (\eta_k^c) \ket_{\Theta} = \sqrt{2} \mathcal{R}[y_k(\eta_k^c) d_{\nk}^{R,I}]=0$, $\bra \hat{\pi}_{\nk}^{R,I} 
%(\eta_k^c) \ket_{\Theta} = \sqrt{2} \mathcal{R}[g_k(\eta_k^c) d_{\nk}^{R,I}]$. Using the latter relations to express $d_{\nk}^{R,I}$ in terms of the expectation values at the time of collapse 
%and substituting it in \eqref{expeceta},
  This  can be  done in various  ways   but the simplest invokes  using Ehrenfest's  theorem  
to obtain  the expectation values of the field operators  at  any later time    in terms of the expectation values at the time of 
collapse. The result is:

\beq\label{expecyeta}
\bra \hat{y}_{\nk}^{R,I} (\eta) \ket_\Theta = \bigg[ \frac{\cos (k\eta - z_k)}{k} \bigg( \frac{1}{k\eta} - \frac{1}{z_k} \bigg) + \frac{\sin (k\eta - z_k)}{k} \bigg(\frac{1}{k\eta z_k} + 1 \bigg) \bigg]  \bra 
\hat{\pi}_{\nk}^{R,I} (\tc) \ket_\Theta,
%&+& \bigg( \cos (k\eta - z_k) - \frac{\sin (k\eta - z_k)}{k\eta} \bigg)  \bra \hat{y}_{\nk}^{R,I} (\tc) \ket_\Theta,
\eeq

\beq\label{expecpieta}
\bra \hat{\pi}_{\nk}^{R,I} (\eta) \ket_\Theta  = \bigg( \cos (k\eta - z_k) +\frac{\sin (k\eta - z_k)}{z_k} \bigg) \bra \hat{\pi}_{\nk}^{R,I} (\tc) \ket_\Theta,  
%- k \sin (k\eta - z_k)  \bra \hat{y}_{\nk}^{R,I} (\tc) \ket_\Theta,
\eeq
with $z_k \equiv k\tc$. This calculation is explicitly done in Refs. \cite{sudarsky2006,adolfo2010}.

Finally, using \eqref{master2},  \eqref{momentito1}  and \eqref{expecpieta} we find and expression for the Newtonian potential in terms of the random variables and the time of collapse 

\beq\label{masterrandom}
\Psi_{\nk} (\eta) =     \frac{ \sqrt{\epsilon}  H}{ M_P} \left( \frac{L}{2k} \right)^{3/2}   \left( \cos (k\eta - z_k)+ \frac{\sin (k\eta - z_k)}{z_k} \right) X_{\nk}, %- X_{\nk}^{(1)} \sin (k\eta - z_k) \left( 1 + \frac{1}{z_k^2} \right)^{1/2}   ,
\eeq
where $X_{\nk} \equiv x_{\nk}^R + i x_{\nk}^I$, and taking into account Eqs \eqref{equisri}, it is clear that

\begin{subequations}\label{promequis}
\beq
\overline{X_{\nk} X_{\nk'}^*}= 2[ \delta_{\nk,\nk'} + \lambda(\delta_{\nk,2\nk'}+\delta_{2\nk,\nk'})],
\eeq
\beq
\overline{X_{\nk} X_{\nk'}}=\overline{X_{\nk}^* X_{\nk'}^*} = 2[ \delta_{\nk,-\nk'} + \lambda(\delta_{\nk,-2\nk'}+\delta_{2\nk,-\nk'})].
\eeq
\end{subequations}

Expression \eqref{masterrandom} is the main result of the present section. It relates the Newtonian potential during inflation to the parameters 
describing the collapse (i.e. the random variables and the time of collapse). It is worth noting that all the quantities occurring  in \eqref{masterrandom} are  all complex valued  quantities and   no quantum operators  appear in the  expression. This is an  important   difference between what one  must   confront  in  our approach and  in the standard treatment  of perturbations  during    inflation. That is,  in 
the latter approach, the Newtonian potential is strictly 
a quantum operator and then   one needs to  invoke various   kinds of arguments that are  often   quite vague,   and  do  not  lead to clearly  defined  connections  with the quantities found in the observations; in particular, one   often    finds  in the    discussions   presented in the  standard approaches to the subject, an  appeal to  quantum  randomness  that is,    however, left   completely unspecified. 
Thus,   the standard   approach   suffers  from the lack  of opportunity  for  completely  clarify the  characterization regarding the stochastic aspects  of the   situation  (as  well  as   from  other  conceptual  deficiencies  that  have  been have discussed  in \cite{sudarsky2009}).
In our approach, we  will not   rely on  arguments  involving horizon-crossing of the modes, decoherence or many worlds interpretation of 
quantum mechanics, 
to justify the transition from a quantum object $\hat \Psi$ to a classical stochastic field $\Psi$, which as we said  often  leads  to rather vague   notions  about the  connection of the  mathematical  expressions used  and  the   objects 
that emerge  from  observations.  One of the  advantages of the approach  we favor is that, as a 
result of the  collapse  postulate, such connection  become   transparent and  specific:   it is fully  
encoded  in  the variables  $X_{\nk}$  characterizing,   in a unequivocal manner,  all and every stochasticity   we  will  
need  to deal with.

As is well known, the Newtonian potential is closely  related with the  the temperature anisotropies  whose origins  can be traced back (in the specific gauge)  with the extra  red/blue shift  photons  suffered  when emerging from the local potential wells/hills. As  the values of the two random 
variables associated to each mode, $x_{\nk}^{R}$ and $x_{\nk}^{I}$, are fixed for our universe, it follows  from  expression \eqref{masterrandom} that  these values determine   the  value  of the Newtonian potential Fourier components  corresponding to  our  universe, which in turn fix the value of the   observed temperature anisotropies.  The statistic nature of  the prescribed  distribution of the random variable $X_{\nk} \equiv x_{\nk}^{R}+ i x_{\nk}^{I} $ gets transfered to the Newtonian potential $\Psi_{\nk}$. It is   clear that we cannot give a definite prediction for the values that  these random variables take in our universe, given the  intrinsic randomness of  the collapse. 
 %(  which  would  broadly correspond  to    the  intrinsic    randomness   associated  with Quantum Mechanics  in  those  situations  involving  a Meassurment prices).
 However,  as  we will show next,   the fact that we have  a large number  of modes  $\vec k$   contributing to each of the observed  quantities, is  the   feature  that allows a statistical  analysis and  the making of {\it a priori } theoretical estimates for the observational  quantities.
%that will be the focus of the next section.

\subsection{Observational quantities}\label{angularspectrum}

In this subsection we will present a brief review of the main results obtained in Ref. \cite{sigma} which  show how the correlations between different modes, produced by the self-induced collapse, affect the observational quantities, in particular, the CMB angular spectrum.  In said work we explored a correlation between the random variables  of any mode  with those of their higher harmonics (which is  in a sense  reminiscent of the  so-called  parametric   resonances  found  in  quantum optics in materials  with  nonlinear   response  functions \cite{Parametric-resonance}, and  just like those has  its origins in the intrinsic  non-linearity of  the  problem at hand ). As mentioned in \cite{sigma}, we found that this effect leads to a departure from the standard prediction afflicting mainly  the first multipoles of the angular power spectrum. 

The observational quantity of interest corresponds to the temperature fluctuations of the CMB observed today on the celestial two-sphere.\footnote{ The two sphere  corresponding to the intersection of our  present day  past light  cone   with  the  last  scattering hypersurface} The temperature anisotropies are expanded using the spherical harmonics $ \frac{\delta T}{T_0} (\theta,\varphi) = \sum_{l,m} a_{lm} Y_{lm} (\theta,\varphi)$, which means that the coefficients $a_{lm}$ can be expressed as

\beq\label{alm0}
a_{lm} = \int \frac{\delta T}{T_0} (\theta,\varphi) Y_{lm}^\star (\theta,\varphi) d\Omega,
 \eeq
here $\theta$ and $\varphi$ are the coordinates on the celestial two-sphere, with $Y_{lm}(\theta,\varphi)$ the spherical harmonics ($l=0,1,2...$ and $-l \leq m \leq l$), and $T_0 \simeq 2.725$ K the temperature average. 

The different multipole numbers $l$ correspond to different angular scales; low $l$ to large scales and high $l$ to small scales. At large angular scales ($l \leq 20$), the Sachs-Wolfe effect is essentially the   single aspect determining  the temperature fluctuations in the CMB. That effect relates the anisotropies in the temperature observed today on the celestial two-sphere to the inhomogeneities in the last scattering surface,

\beq
\frac{\delta T}{T_0} (\theta,\varphi) \simeq \frac{1}{3} \Psi^{\text{matt}} (\eta_D, \x_D),
\eeq
where $\x_D= R_D (\sin \theta \sin \varphi, \sin \theta \cos \varphi, \cos \theta)$, with $R_D$ the radius of the last scattering surface and $\eta_D$ is the conformal time of decoupling ($R_D \simeq 4000$ Mpc, $\eta_D \simeq 100$ Mpc). The Newtonian potential can be expanded in Fourier modes, $\Psi^{\text{matt}} (\eta_D, \x_D) = \sum_{\nk} \Psi_{\nk}^{\text{matt}} (\eta_D) e^{i \nk \cdot \x_D} / L^3$ . Furthermore, using that $e^{i \nk \cdot \x_D} = 4 \pi  \sum_{lm} i^l j_l (kR_D) Y_{lm} (\theta,\varphi) Y_{lm}^\star (\hat k )$,  expression \eqref{alm0} can be rewritten as

\beq\label{alm1}
a_{lm} = \frac{4 \pi i^l}{3L^3} \sum_{\nk} j_l (kR_D) Y_{lm}^\star(\hat k) \Psi_{\nk}^{\text{matt}} (\eta_D),
\eeq
with $j_l (kR_D)$ the spherical Bessel function of order $l$.

The Newtonian potential $\Psi^{\text{matt}}$ appearing in \eqref{alm1} is evaluated at the time of decoupling which corresponds to the matter dominated cosmological epoch. Traditionally, the relation between $\Psi^{\text{matt}}$ and the Newtonian potential at the end of inflation is made by making use of the so-called transfer functions $T(k)$; the transfer functions 
contain all relevant physics from the end of inflation to the latter matter dominated epoch, which includes among others the acoustic oscillations of the plasma, and  are the  source of the   main modifications of the simple  scale invariant  H-Z  spectra. Thus, $\Psi_{\nk}^{\text{matt}} (\eta_D) = T(k) \Psi_{\nk}$,  where $\Psi_k$ corresponds to the Newtonian potential during inflation and, since one is interested in the modes with scales of observational interest,  the modes $\Psi_{\nk} (\eta)$ must satisfy $-k\eta \ll 1$ (or, as commonly referred, its scale should be ``well outside the horizon'' during inflation).  This is, the coefficients $a_{lm}$ must be  rewritten as

\beq\label{alm2}
a_{lm} = \frac{4 \pi i^l}{3L^3} \sum_{\nk} j_l (kR_D) Y_{lm}^\star(\hat k) T(k) \Psi_{\nk}.
\eeq

At this point the traditional approach would proceed to calculate averages and higher-correlation 
functions of the coefficients $a_{lm}$. Nevertheless, within our model we can make a further step, by 
substituting Eq. \eqref{masterrandom} (and taking the limit $-\eta \to 0$), which gives an explicit 
expression for $\Psi_{\nk}$ in terms of the parameters of the collapse, in Eq. \eqref{alm2}, one 
obtains\footnote{Note that we have multiplied by a factor of $3/(5 \epsilon)$ the $\Psi_{\nk}$ we 
obtained during inflation, Eq. \eqref{masterrandom}. This is because, while $\Psi_{\nk} (\eta)$ is 
constant for modes $-k\eta \ll 1$ during any cosmological epoch, its behavior changes substantially 
during a change in the equation of state for the dominant type of matter in the universe. In particular, 
during the change from inflation to radiation epochs, $\Psi$ is amplified by a factor of approximately 
$1/\epsilon$. For a detailed discussion regarding the amplitude within the collapse framework see Ref. 
\cite{gabriel2010}. 
 }

\beq\label{almrandom}
a_{lm} = \frac{i^l}{L^{3/2}} \sum_{\nk} g(z_k) F_{lm} (\nk) X_{\nk},
\eeq
with 

\beq\label{Flm}
F_{lm} (\nk) \equiv   \frac{ j_l (kR_D) Y_{lm}^{\star} (\hat k)}{k^{3/2}} T(k),
\eeq
and

\beq\label{gz}
g(z_k) \equiv  \frac{\pi H}{5 M_P} \sqrt{\frac{2}{\epsilon}} \bigg( \cos z_k   - \frac{\sin z_k}{z_k} \bigg).
\eeq

Equation \eqref{almrandom} allow us to appreciate one of the advantages of the collapse proposal: the coefficient $a_{lm}$, which is directly associated with the observational quantities (i.e. the temperature fluctuations), is in turn related to the random variables characterizing the collapse. In other words, the statistical features of the coefficients $a_{lm}$ can 
be discussed   in terms of  the statistics of the random variables $X_{\nk}$. 
% The intrinsic randomness of these variables is inherited by the quantum theory, and when considering the statistical 
%aspects of the random variables, we will be able to make a prediction for the observational quantities. 
We note   that  there is no analog expression of Eq. \eqref{almrandom} in the 
standard approach. As a matter of fact, if we follow the conventional way of identifying quantum expectation values with classical quantities, the prediction given by the standard inflationary paradigm would be $\bra 0 | \hat \Psi_k |0\ket = 0 = \Psi_k$; thus, we  would be lead, by Eq. \eqref{alm2}, to conclude that
 $a_{lm}=0$; this is, the theoretical prediction for the temperature fluctuations would be exactly zero in an evident contradiction\footnote{Several kinds of  arguments would normally  be invoked  at this point  in  defense of the standard treatments.  For a  detailed discussion of their merits and  shortcomings  see Ref. \cite{sudarsky2009}.} with the observations (see Ref. \cite{susana2013}).

Furthermore, from Eq. \eqref{almrandom}, we can calculate the average over possible outcomes of the random variables of the product of two $a_{lm}$ coefficients, this is,

\beq\label{alm2ptos0}
\overline{a_{lm} a_{l'm'}} = \frac{i^{l+l'}}{L^3} \sum_{\nk,\nk'} g(z_{k}) g(z_{k'})F_{lm} (\nk) F_{lm} (\nk') \overline{X_{\nk} X_{\nk'}},
\eeq
then using Eqs. \eqref{promequis}

\bea\label{alm2ptos1}
\overline{a_{lm} a_{l'm'}} &=& \frac{2i^{l+l'}}{L^3} \sum_{\nk,\nk'} g(z_{k}) g(z_{k'})F_{lm} (\nk) F_{lm} (\nk') [ \delta_{\nk,-\nk'} + \lambda(\delta_{\nk,-2\nk'}+\delta_{2\nk,-\nk'})] \nonumber \\
&=& \frac{2i^{l+l'}}{L^3} \sum_{\nk}  g(z_k)^2 \left\{ F_{lm} (\nk) F_{l'm'} (-\nk) + \lambda [F_{lm} (\nk) F_{l'm'} (-2\nk) +  F_{lm} (\nk) F_{l'm'} (-\nk/2)        ] \right\}  \nonumber \\
&=& \frac{2i^{l+l'} (-1)^{l'}}{(2\pi)^3} \int d^3 k \enskip g(z_k)^2 \left\{ F_{lm} (\nk) F_{l'm'} (\nk) + \lambda [F_{lm} (\nk) F_{l'm'} (2\nk) +  F_{lm} (\nk) F_{l'm'} (\nk/2)        ] \right\}, 
\eea
where in the last line we used  parity of the spherical harmonics, which implies that $F_{lm} (-\nk) = (-1)^l F_{lm} (\nk)$ and took the limit $L \to \infty$ and $\nk$ continuous. As we have found in previous works, the data indicates that, to a good degree of accuracy \cite{adolfo,susana2012}, the assumption that $z_{k} \equiv k\tc$ is independent of $k$, i.e. the time of collapse goes as $\tc \propto k^{-1}$. Therefore, assuming $z_k = z =$ const. and taking $T(k)=1$, which  as   previously  indicated        is a  very  good approximation for $l \leq 20$, one can compute the integrals in Eq. \eqref{alm2ptos1}, this is, 

\beq\label{alm2ptosa}
\overline{a_{lm} a_{l'm'}} = (-1)^m \delta_{ll'} \delta_{m-m'} D_l,
\eeq
with 

\beq\label{Dl}
D_l \equiv \frac{g(z)^2}{(2\pi)^3} \frac{1}{l(l+1)} \left[ 1 + \lambda G(l) \right];
\eeq
we have also defined $G(l)$ as 

\beq\label{Gl}
G(l) \equiv \frac{9l(l+1)\sqrt{\pi} \Gamma (l) }{2^{l+5/2} \Gamma(l+3/2)} \quad {}_2F_1 (l,-1/2;l+3/2,1/4),
\eeq
where $\Gamma(l)$ is the Gamma function and $_2F_1 (\alpha,\beta; \gamma,z)$ is the Hypergeometric function.\footnote{The Hypergeometric function is defined as $_2F_1 (\alpha,\beta; \gamma,z) \equiv \frac{\Gamma(\gamma)}{\Gamma(\beta) \Gamma(\gamma- \beta)} \int_0^1 \frac{t^{\beta-1} (1-t)^{\gamma - \beta-1}}{(1-tz)^\alpha} dt$.}

 %Similar calculation gives
 
%  \begin{subequations}
%  \beq\label{alm2ptosb}
%  \overline{a_{lm} a_{l'm'}} = \overline{a_{lm}^* a_{l'm'}^*}, 
%  \eeq
%  \beq\label{alm2ptosc}
%  \overline{a_{lm} a_{l'm'}^*} = \delta_{ll'} \delta_{mm'} D_l .
%  \eeq
% \end{subequations} 

%[ D: REDUNDANTE] 

As  we  said,  a   key   aspect that  in  our  treatment  differs,  from those followed  in the   standard   approaches,  
is  the manner in which  the  results  from the formalism  are   connected  to observations.  This  is    
most clearly  exhibited  by   our result regarding the quantity  $a_{lm} $   in Eq. \eqref {almrandom}.  
Despite the fact that we have in principle  a close    expression for   the quantity of interest,  we  
cannot use Eq.  \eqref {almrandom}  to make  a definite prediction  because   the  expression involves 
the  numbers   $ X_{\nk}$  that correspond, as  we  indicated  before,  to   a    random    choice  
``made  by  nature''   in the context of the collapse process. 
 The  way one  make  predictions  is by noting  that   the  sum  appearing   in Eq.   \eqref {almrandom}      
 represents a   kind of  ``two-dimensional  random walk,''  i.e. the sum   of complex numbers   
 depending on    random  choices  (characterized by the $ X_{\nk}$).  As  is  well known, for a random  
 walk,  one cannot   predict the final displacement (which  would correspond to the   complex  quantity 
 $a_{lm} $),    but  one  might estimate  the  most likely value of the  magnitude  of such  
 displacement.  Thus,  we focus  precisely  on the  most likely  value of $|a_{lm}|$,  which we  denote  
 by   $|a_{lm}|_{\text{M.L.}}$.    In order to compute  that  quantity,  we   make  use  of a   fiducial  
 (imaginary)  ensemble of  realizations of the random  walk  and  compute the   average over the 
 ensemble    of the    value   of  the total displacement ($\overline{|a_{lm}|}$).  Finally   we   identify:
 
 \beq\label{ML}
|a_{lm}|_{\text{M.L.}} \simeq \overline{|a_{lm}|}.
\eeq

%The overline appearing denotes average over the    fiducial   ensemble  of realizations,    which  would correspond  to  an  imaginary ``ensemble of universes.'' 
Thus, using the previous result, Eq. \eqref{alm2ptosa} (with $l=l'$ and $m=m'$), we  find 

\beq
|a_{lm}|_{\text{M.L.}} \simeq \overline{|a_{lm}|^2} = \overline{a_{lm} a_{lm}^*} = D_l.
\eeq

  At   this  point, one   could  focus   on   the   quantity  that   is  most  often studied  in this context, namely
     
   \beq\label{ML5}
 C_l  \equiv \frac{1}{2l+1}  \sum_m  |a_{lm}|^2,
    \eeq
 for which  we  would  have the estimate
      \beq\label{ML6}
 {C_l}^{\text{M.L.}}  \equiv \frac{1}{2l+1}  \sum_m  |a_{lm}|_{\text{M.L.}}^2 \simeq D_l.
    \eeq
    This is, the angular power spectrum is given by $l(l+1) C_l^{\text{M.L.}} =  l(l+1)D_l$, explicitly:

\beq\label{clML}
l(l+1)C_l^{\text{M.L.}} = \frac{g(z)^2}{(2\pi)^3} \left[ 1 + \lambda G(l) \right].
\eeq
It is clear that  if $\lambda = 0$, then the probability distribution functions  for the variables 
$X_{\nk}, X_{\nk'}$ are  uncorrelated,  random, and Gaussian (see \eqref{equisr} and \eqref{equisi}. 
Furhtermore, for $\lambda=0$ we recover standard flat spectrum (recall that we are assuming that 
$z=k\tc$ is independent of $k$), i.e. $l(l+1)C_l$ is independent of $l$ [see Eq. \eqref{clML}].  The 
departure of the flat spectrum has a very specific  signature that, in principle, can be searched for  
observationally. The behavior of $G(l)$ for  $2 \leq l \leq 20$ is shown in Fig. 1.

     \begin{figure}[htbp]
      \begin{centering}
      \includegraphics{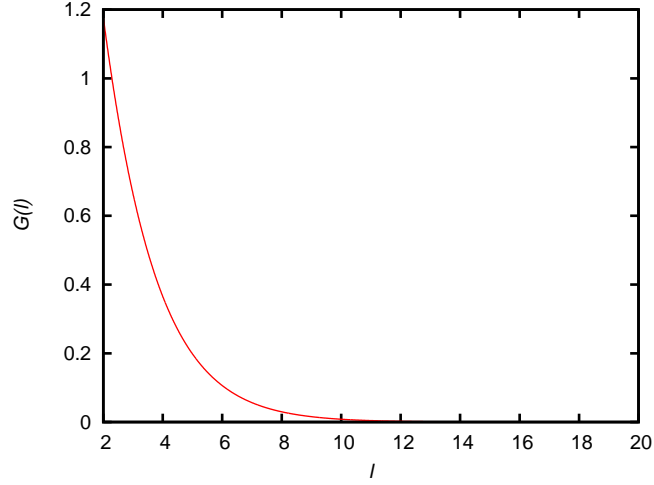}
      \label{graf1}
      \end{centering}
      \caption{Plot of the function $G(l)$ [defined in \eqref{Gl}] in the range $2\leq l \leq 20$. The function $G(l)$, which encodes the correlation between the modes induced by the collapse, affects only the lowest multipoles}
     \end{figure}

Clearly, by ignoring the transfer function, the traditional prediction of the angular power spectrum is a constant; on the other hand, in our approach, the mode correlation induced by the self-induced collapse, altered the standard prediction by adding the term $\lambda G(l)$. In Fig. (2a), we show a graphic of our predicted value for $l(l+1)C_l$ [normalized see Eq. \eqref{clML}]  with three different values of the $\lambda$ parameter and also a best-fit constant (799.25) to the data from \emph{Planck} collaboration.\footnote{See \url{http://pla.esac.esa.int/pla/\#cosmology}} In Fig. (2b) we show a plot for our predicted angular spectrum with the best-fit value $\lambda=-0.3039$ to \emph{Planck} data. We see that our prediction is, in principle, able to improve the theoretical estimate to the observational data. 

\begin{figure}
 \centering
\includegraphics[width=7.5cm,height=6.0cm]{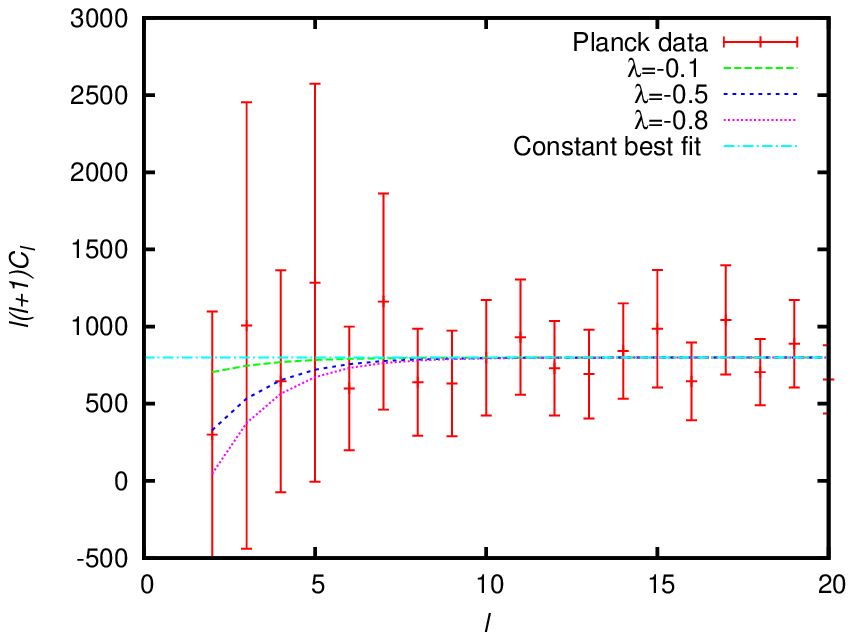}
\includegraphics[width=7.5cm,height=6.0cm]{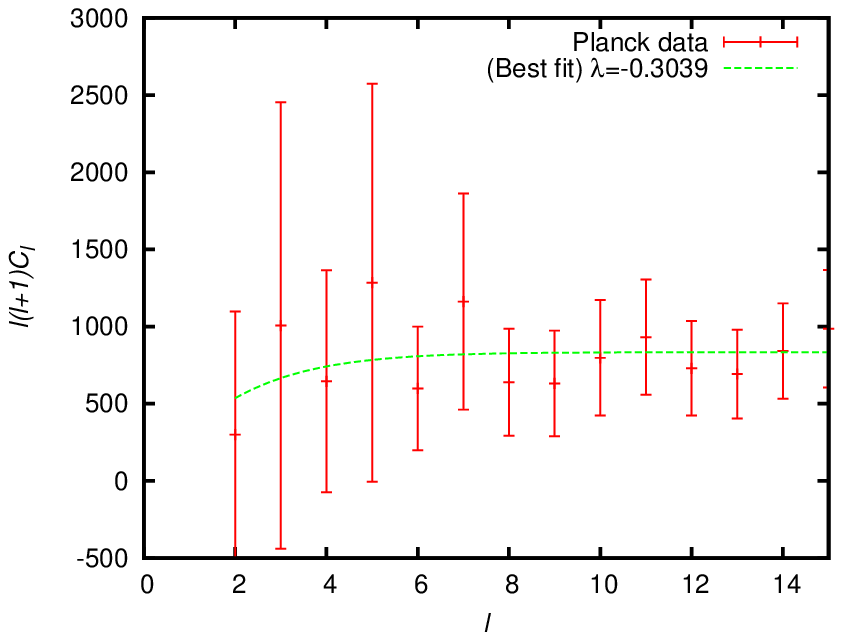}
 \caption{Fig. (2a) shows our predicted value for $l(l+1)C_l$ [Eq. \eqref{clML} normalized] in the range $2\leq l \leq 20$ considering three values of $\lambda$ and the best-fit constant to the observational data from \emph{Planck} collaboration. In Fig. (2b) we show our prediction for $l(l+1)C_l$ [Eq. \eqref{clML} normalized] in the range $2\leq l \leq 15$ with the best-fit value $\lambda=-0.3039$ to \emph{Planck} data.     }
 \label{f:graficos2}
\end{figure}

As we can see, the effect is stronger for low $l$ and it decreases in a nearly exponential fashion for 
large $l$. For a more complete analysis, we should take into account the transfer functions $T(k)$ 
that are associated with the acoustic oscillations originated after the end of the inflationary era, this 
would lead to the full angular power spectrum. We want to emphasize that the result in Eq. \eqref{clML} is different from the standard approach; first, it contains the function $g(z)$, which is a  function of the time of collapse, and second, the correlation between different modes provided by the 
collapse is encoded in the function $G(l)$. Evidently, a more detailed analysis should be performed before reaching any conclusions, but this modest analysis shows the promising potential of our approach.

%     
%    where in the last  step,   we  used the fact that   the     most  likely value   estimate  in Eq.  \eqref{ML4}  is  independent of  $m$. Furhtermore, if we consider the time of collapse as $\tc \propto k^{-1}$, i.e. $z_k = z$ independent of $k$ and take $T(k)=1$, which is a valid approximation for $l \ll 20$,  we recover an exact scale-invariant spectrum, this is, 
%    
%    \beq\label{CL}
%    l(l+1) C_l^{\text{M.L.}} = \frac{g(z)^2}{(2\pi)^3} = \frac{H^2}{10^2 \pi M_P^2 \epsilon } \left( \cos z - \frac{\sin z}{z} \right)^2 \equiv A
%    \eeq
%    The quantity $A$ is fixed by the observational data to be $A \simeq 10^{-10}$. The fact that $\tc \propto k^{-1}$ is also motivated by the results in  previous works \cite{sudarsky2006,gabriel2010,adolfo,adolfo2010,alberto2012}  
% 
%    
%    If we would like to recover the full angular spectrum, one should then use expression \eqref{ML6} including the transfer functions, which can be obtained using numerical codes, and assume a particular form for the time of collapse $\tc$ in terms of $k$. This type of studies have been done, and the results can be consulted in Ref. \cite{susana2012}. 

Expression \eqref{clML} is the  theoretical estimate  to be compared  with the observational    data, 
and  as   should be clear from  the discussion,  the fact that we  have to rely on   most likely  values,  
for what  are    in  effect    the  mathematical equivalent of random walks,  leads us to     expect  that 
there  should  be  a   general and rough  agreement    between our estimates and  observations 
(assuming the theory is  correct). However, we  do not really   expect    a    detailed  and precise  
match    simply  due to the  intrinsic  randomness   involved  and  characterized  here  in full   in terms 
of the random numbers  $X$. In the  standard  approach,    similar considerations involving the  
randomness of the fluctuations  and the   uncertainties   tied to  stochasticity,     mixed together   
with   arguments   involving the limited     nature of  the region of the universe  one  is  observing,   
also lead to people in the community  to   expect   small  differences in predictions  and observations. 
Nevertheless, in general, such   discussions  are based on heuristic   arguments;  therefore,  are  
limited   both in scope and precision.  The  essential   difficulty is that, in the standard  analysis,  the    
precise  stochastic  elements  are not  identified and have no  mathematical   representation in the  
formalism.   We  believe  that,  the fact  that  in our  approach,   the stochastic  elements    are  
clearly identifiable    (i.e.  the  $ X_{\nk}$),   represents  a great  advantage.   For instance,    the  
formalism  here   provides us with  an explicit expression for the  quantity $ a_{lm}  $,  such as  in  Eq. 
\eqref{almrandom},  and, thus,  allowing us to study in great detail and  in a transparent manner,    
the  precise  nature of   higher  
order statistical   estimates    as  we  will  do  in the following.

\section{Characterizing the primordial CMB bispectrum}\label{seccolapsobispec}

In this section, we will study the connection between the Newtonian potential at the end of inflation and the observational quantities obtained from the temperature anisotropies 
in the CMB; in particular, we will provide the connection between the  parameters characterizing the collapse and the primordial \textit{bispectrum}.
 %The goal of the entire section will be to attain a prediction for the  bispectrum when considering the 
%collapse proposal.

%\subsection{The primordial bispectrum}\label{colapsobispec}

The usual path to look for  non-trivial statistical features (e.g. possible non-Gaussianities) in the CMB is to study the {bispectrum}, which is  considered  to be directly related with the three-point function of the temperature anisotropies in harmonic space. The theoretical CMB angular bispectrum is defined as

\beq\label{bispec0}
B^{l_1 l_2 l_3}_{m_1 m_2 m_3} \equiv \overline{a_{1_1 m_1} a_{1_2 m_2} a_{1_3 m_3}}.
\eeq

The bar appearing in \eqref{bispec0} denotes average over an ensemble of universes, but in practice such averaging  is not  viable and  is  usually  ``replaced,'' in a sense,  by  averaging over orientations in our own universe; the relation  between the two types of averages is not  clear or direct (this fact has been discussed in   great detail  in Ref. \cite{susana2013}). In the following, we will show how our approach helps to 
clarify certain  issues  that   emerge when dealing with the statistical aspects of the spectrum and  
when comparing  theoretical   estimates  and    observations.

Given the definition of the CMB bispectrum and  by  using considerations   involving  the assumption of  an (in average) rotational invariant sky,\footnote{ Here we can   see some of the difficulties   encountered in  standard discussions: If one  considers the   average   over universes,  one  must face  not only the fact  that  this  ensemble,  even if it  is   real,   is  empirically   inaccessible, and if it is  over orientations then  the assumption of rotational invariance is  simply empty.  In  fact, if the   ensemble  is  just an  imaginary ensemble   the rotational   invariance  can be    obtained  by construction starting with  any  arbitrary  universe and  constructing the ensemble   by  including all possible rotations of the  original one.  It is  clear that such  considerations   have no bearing   on the  properties of the  universe  we have   access to  with our observations. } one finds in the literature \cite{komatsu2003,komatsu2009} another object called the ``angle-averaged bispectrum'' defined by

\beq\label{avg}
B_{l_1 l_2 l_3} \equiv \sum_{m_i}  \left( \begin{array}{lcr}
      l_1& l_2 & l_3  \\
     m_1 & m_2 & m_3
    \end{array}
    \right) B^{l_1 l_2 l_3}_{m_1 m_2 m_3} = \sum_{m_i}  \left( \begin{array}{lcr}
      l_1& l_2 & l_3  \\
     m_1 & m_2 & m_3
    \end{array}
    \right) \overline{a_{1_1 m_1} a_{1_2 m_2} a_{1_3 m_3}}.
\eeq
The object $ \left( \begin{array}{lcr}
      l_1& l_2 & l_3  \\
     m_1 & m_2 & m_3
    \end{array}
    \right) $ is called the Wigner 3-$j$ symbol (see Ref. \cite{Komatsu2001} for more details and properties for these functions); the bispectrum $B_{l_1 l_2 l_3}$ is non-vanishing for the values of  $l,m$  satisfying  the following conditions:

\begin{enumerate}
 \item $m_1 + m_2 + m_3 = 0$.
 
 \item $ l_1+l_2 + l_3$ is even. 
 
 \item  $|l_i-l_j| \leq l_k \leq l_i +l_j$ for all permutations of indices.
 
\end{enumerate}
 Here it is important to mention that the 2nd. selection rule, comes not from the selection rules corresponding to the Wigner 3-j symbols, but from the additional assumption that the 3-point correlation function of the temperature anisotropies must be invariant under spatial rotations and translations \cite{Luo1994,Gangui1999,Gangui2000}. This is a crucial difference with our approach as we will see in the following. Also, the previous three selection rules are called ``the triangle conditions'' as $l_1,l_2,l_3$ must correspond to the sides of a triangle. As a matter of fact, in the standard approach, one intents to estimate $B_{l_1 l_2 l_3}$ from the observational data (e.g. see  Sec. 3.1  of Ref. \cite{planckng}) by testing different configurations for such ``triangles.''

Motivated by the fact that, when  following  our  approach,  we can obtain a direct relation between the coefficients $a_{lm}$ and the random variables characterizing the collapse [Eq. \eqref{almrandom}] we will be  focussing  on the  expression for   the  ``observational"   bispectrum:  

\beq\label{bispecorig}
\mathcal{B}^{\text{obs}}_{l_1 l_2 l_3} \equiv \sum_{m_i}  \left( \begin{array}{lcr}
      l_1& l_2 & l_3  \\
     m_1 & m_2 & m_3
    \end{array}
    \right) a_{1_1 m_1} a_{1_2 m_2} a_{1_3 m_3} .
\eeq
%and from now on we will refer to this object as the ``collapse bispectrum.''
 Note that this object also contains the Wigner 3-$j$ symbol, therefore, $l_1,l_2,l_3$ must satisfy the first and the third selection rule, meanwhile the 2nd. rule needs not to be satisfied, since the observed bispectrum is certainly not rotational nor translational invariant; the symmetry associated to the 2nd. selection rule applies to the average not to the observed bispectrum which is just one element (realization) of the ``ensemble.'' As a matter of fact, the more general rule, which encompass the 2nd. selection rule, for the Wigner 3-$j$ symbol is $l_1+l_2+l_3=$ integer (or even if $m_1=m_2=m_3=0$).

The difference between $B_{l_1 l_2 l_3}$ and $\mathcal{B}^{\text{obs}}_{l_1 l_2 l_3}$ is a subtle but important one. While in the definition of $B_{l_1 l_2 l_3}$ one  should  perform  an average over an ensemble of universes [as is explicitly stated in the definition \eqref{avg}], the object $\mathcal{B}^{\text{obs}}_{l_1 l_2 l_3}$   involves no averages over idealized  ensembles whatsoever.\footnote{One  should  not confuse  the  fact that when  obtaining the  specific  values of $ a_{lm}$, which  result from observations,   one  needs to perform an integral over   the CMB sky [as  indicated in  Eq. \eqref{alm0}],  with taking averages  over  ensembles of universes  as considered   above.} The only average that is being performed in $\mathcal{B}^{\text{obs}}_{l_1 l_2 l_3}$, is an average over orientations (i.e. a sum over $m_i$ with a weight given by the Wigner 3-$j$ symbols).

In our approach, the prediction for $\mathcal{B}^{\text{obs}}_{l_1 l_2 l_3}$ is given explicitly by substituting \eqref{almrandom} in \eqref{bispecorig} which yields

\beq\label{bispecorig2}
\mathcal{B}^{\text{obs}}_{l_1 l_2 l_3} = \sum_{m_i}  \left( \begin{array}{lcr}
      l_1& l_2 & l_3  \\
     m_1 & m_2 & m_3
    \end{array}
    \right) \frac{ i^{l_1+l_2+l_3} }{L^{9/2}} \sum_{\nk_1, \nk_2, \nk_3} g(z_{k_1}) g(z_{k_2}) g(z_{k_3}) F_{l_1 m_1} (\nk_1) F_{l_2 m_2} (\nk_2) F_{l_3 m_3} (\nk_3) X_{\nk_1} X_{\nk_2} X_{\nk_3}.
\eeq
% \bea\label{bispecorig2}
% \mathcal{B}^{\text{obs}}_{l_1 l_2 l_3} &=& \sum_{m_i}  \left( \begin{array}{lcr}
%       l_1& l_2 & l_3  \\
%      m_1 & m_2 & m_3
%     \end{array}
%     \right) \frac{ 1 }{L^{9/2}} \sum_{\nk_1, \nk_2, \nk_3} \frac{g(z_{k_1}) g(z_{k_2}) g(z_{k_3}) j_{l_1}(k_1 R_D) j_{l_2}(k_2 R_D) j_{l_3}(k_3 R_D) }{(k_1 k_2 k_3 )^{3/2}} \nonumber \\
% &\times& Y_{l_1 m_1}^\star (\hat{k}_1) Y_{l_2 m_2}^\star (\hat{k}_2) Y_{l_3 m_3}^\star (\hat{k}_3) T(k_1) T(k_2) T(k_3)  X_{\nk_1} X_{\nk_2} X_{\nk_3}.
% \eea
As is clear from Eq. \eqref{bispecorig2}, the collapse bispectrum is in effect a sum  of  random complex  numbers (i.e. a sum where each term is characterized by the product  $ X_{\nk_1} X_{\nk_2} X_{\nk_3}$, which is itself,   a complex random number), leading to what can be considered effectively as a two-dimensional (i.e. a  complex plane)  random walk,    in complete  analogy  with   what  we  found in dealing with our estimates of  $a_{lm}$  in \eqref{almrandom}.  Once more, one 
cannot give a perfect estimate for the  final displacement  resulting from the random walk. 
Nevertheless, one might give an estimate of its magnitude. 
%It is for that reason that we will be 
%focusing on estimating the most likely value of the magnitude  $| \mathcal{B}_{l_1 l_2 l_3}|^2$. 

Similarly   as  $\mathcal{B}^{\text{obs}}_{l_1 l_2 l_3}$ is characterized by the  sum  of  random variables
% introduced by the collapse hypothesis, 
we cannot give a specific value for its outcome.  However,  as   we  will see, and   in complete analogy  of  our  analysis of   the quantities $a_{lm}$, by focusing on the most likely value of the magnitude  $| \mathcal{B}^{\text{obs}}_{l_1 l_2 l_3}|^2$  we  will obtain  a reasonable  prediction. 
% In order to be able to make predictions we need further considerations.

To  recapitulate,  the original  situation   corresponds  to the homogeneous and isotropic  vacuum state. When a sudden change of the 
initial state takes place due to  the collapse  (one for each mode),  the   mode    becomes characterized   by  a  fixed   value of  the corresponding  random variables; the collection of all the  values  of  such  random variables associated to all the modes characterizes, therefore,  our  single  and unique universe (so  such  numbers  in consequence completely determine  $|\mathcal{B}^{\text{obs}}_{l_1 l_2 l_3}|^2$); let us denote this set by
 
\beq\label{u}
 U = \{  X_{\nk},  X_{\nk'}, \ldots \}.
\eeq
Nevertheless, given the stochastic nature of the collapse,  we can consider that  the universe  could  have corresponded  to  different set of  values  for the  random variables characterizing the universe in a different manner $\tilde{U} = \{ \tilde{X}_{\nk},  \tilde{X}_{\nk'}, \ldots \}$. The collection of different sets $\{ U, \tilde U, \ldots \}$  thus  describe an hypothetical ensemble of universes. We will denote the bispectrum associated to the hypothetical ensemble of universes as

\beq\label{bispecorig2a}
\mathcal{B}_{l_1 l_2 l_3} (\xi) = \sum_{m_i}  \left( \begin{array}{lcr}
      l_1& l_2 & l_3  \\
     m_1 & m_2 & m_3
    \end{array}
    \right) \frac{ i^{l_1+l_2+l_3} }{L^{9/2}} \sum_{\nk_1, \nk_2, \nk_3} g(z_{k_1}) g(z_{k_2}) g(z_{k_3}) F_{l_1 m_1} (\nk_1) F_{l_2 m_2} (\nk_2) F_{l_3 m_3} (\nk_3) X_{\nk_1} (\xi)  X_{\nk_2} (\xi) X_{\nk_3} (\xi),
\eeq
with the letter $\xi$ labeling the possible set of outcomes corresponding to the random variables that characterizes a particular Universe, thus, for example $\xi=\text{obs}$, corresponds to the bispectrum observed in our universe, this is Eq. \eqref{bispecorig2}, which we regard as typical member  of this hypothetical ensemble. 

Furthermore, we will (just  as when dealing with $a_{lm}$),  make the assumption that the most likely (M.L.) value of the magnitude $|\mathcal{B}_{l_1 l_2 l_3} (\xi)|^2$ in such ensemble  is  a good  estimate  of the   corresponding one for our own universe, that is 

\beq
 |\mathcal{B}^{\text{obs}}_{l_1 l_2 l_3}|^2 \simeq |\mathcal{B}_{l_1 l_2 l_3} (\xi)|^2_{\textrm{M.L.}} 	
\eeq

Moreover,  we  can  simplify the estimate   by   taking the ensemble average $\overline{|\mathcal{B}_{l_1 l_2 l_3} (\xi) |^2  } $, i.e. the average over all possible outcomes of the bispectrum labeled by $\xi$ [Eq. \eqref{bispecorig2a}] and identify it with the most likely $|\mathcal{B}_{l_1 l_2 l_3} (\xi)|^2_{\textrm{M.L.}}$. It is needless to say that these two notions are not exactly the same for arbitrary  
kinds of ensembles; as a matter of fact, the relation between the two concepts depends on the probability distribution function (PDF) of the random variables. In principle, we do not 
know the exact PDF, as we have only access to a single realization--our own universe--,  but  as can bee seen from Eqs. \eqref{promequis}, a very good approximation can be made by assuming that the PDF is highly Gaussian, this is, the parameter $\lambda$ which encodes the correlation between the modes caused by the collapse, is small enough that a Gaussian PDF for the random variables is justified. Therefore, we can approximate

\beq
|\mathcal{B}_{l_1 l_2 l_3} (\xi) |^2_{\textrm{M. L.}} \simeq \overline{|\mathcal{B}_{l_1 l_2 l_3}(\xi)|^2 },
\eeq 
%[D:  Aca  hay algo que no me gustaen cuanto a   notacion ...  pues  dijimos  que  ``ObS"  no tiene  nada  que  ver con hacer promedios sobre  ensambles  de unversos  pero el lado derecho   incluye Obs  y   justamenete ese tipo de promedios...    cero   que hay  que  acalarar y cambiar la notacion  ]

which implies that

\beq
|\mathcal{B}^{\text{obs}}_{l_1 l_2 l_3}|^2 \simeq |\mathcal{B}_{l_1 l_2 l_3} (\xi)|^2_{\textrm{M. L.}} \simeq \overline{|\mathcal{B}_{l_1 l_2 l_3} (\xi)|^2}.
\eeq

In the reminder  of this section, we will focus on computing $\overline{|\mathcal{B}_{l_1 l_2 l_3} (\xi)|^2 }$ and will omit the label $\xi$ with the understanding that it represents the ensemble over possible outcomes of $\xi$. Therefore, the ensemble average is

\beq\label{promedios}
\overline{|\mathcal{B}_{l_1 l_2 l_3} |^2} =  \sum_{m_1,\ldots,m_6} \left( \begin{array}{lcr}
      l_1& l_2 & l_3  \\
     m_1 & m_2 & m_3
    \end{array}
    \right) \left( \begin{array}{lcr}
      l_1& l_2 & l_3  \\
     m_4 & m_5 & m_6
    \end{array}
    \right) \overline{a_{l_1 m_1} a_{l_2 m_2} a_{l_3 m_3} a_{l_1 m_4}^\star a_{l_2 m_5}^\star a_{l_3 m_6}^\star}.
    \eeq

% Considering a Gaussian PDF for the random variables $X_{\nk}$ implies taking a Gaussian PDF for $x^R_{\nk}$ and $x^I_{\nk}$ (i.e. the real an imaginary parts of the complex random number $X_{\nk}$). This is, the ensemble average of the products $\overline{x_{\nk}^R x_{\nk'}^R}$ and $\overline{x_{\nk}^I x_{\nk'}^I}$ is characterized by
% 
% \beq\label{promedios}
% \overline{x_{\nk}^R x_{\nk'}^R} = \delta_{\nk,\nk'} + \delta_{\nk,-\nk'}, \qquad \overline{x_{\nk}^I x_{\nk'}^I} = \delta_{\nk,\nk'} - \delta_{\nk,-\nk'}.
% \eeq
% Note that we have taken into account that the variables $x^R_{\nk}$ and $x^I_{\nk}$ are independent. Additionally, we have considered the correlation between the modes $\nk$ and $-\nk$ in accordance with the commutation relation given by $[\hat{a}^{R}_{\nk},\hat{a}^{R \dag}_{\nk'}]$ and $[\hat{a}^{I}_{\nk},\hat{a}^{I \dag}_{\nk'}]$ [see Eq. \eqref{creanRI}]. Given the relations \eqref{promedios}, the average for the product of two random variables $X_{\nk}$ (over the imaginary ensemble of universes) yields
% 
% \beq\label{delta1}
% \overline{X_{\nk} X_{\nk'}} = \overline{(x^R_{\nk} + i x^I_{\nk})(x^R_{\nk'} + i x^I_{\nk'})} = 2 \delta_{\nk,-\nk'}.
% \eeq
% Furthermore, it is easy to check that 
% 
% \beq\label{delta2}
% \overline{X^\star_{\nk} X^\star_{\nk'}} = \overline{X_{\nk} X_{\nk'}} = 2 \delta_{\nk,-\nk'} \qquad \textrm{and} \qquad \overline{X_{\nk} X^\star_{\nk'}} = 2 \delta_{\nk,\nk'}.
% \eeq

Let us focus on the quantity  $\almsix$, thus using Eq. \eqref{almrandom} gives 
\bea\label{alm6ptos}
\almsix &=& \frac{1}{L^6} \sum_{\nk_1, \ldots, \nk_6} g(z_{k_1}) g(z_{k_2}) g(z_{k_3}) g(z_{k_4}) g(z_{k_5}) g(z_{k_6}) \nonumber \\
&\times& F_{l_1 m_1} (\nk_1) F_{l_2 m_2} (\nk_2) F_{l_3 m_3} (\nk_3)    F_{l_1 m_4}^* (\nk_4) F_{l_2 m_5}^* (\nk_5) F_{l_3 m_6}^* (\nk_6) \nonumber\\
&\times& \overline{X_{\nk_1} X_{\nk_2} X_{\nk_3} X_{\nk_4}^* X_{\nk_5}^* X_{\nk_6}^*}.
    \eea% 
% \bea\label{bismodavg}
% \overline{|\mathcal{B}^{\text{obs}}_{l_1 l_2 l_3}|^2} &=&  \frac{ g(z)^6 }{L^9} \sum_{m_1, \ldots, m_6}  \left( \begin{array}{lcr}
%       l_1& l_2 & l_3  \\
%      m_1 & m_2 & m_3
%     \end{array}
%     \right)  \left( \begin{array}{lcr}
%       l_1& l_2 & l_3  \\
%      m_4 & m_5 & m_6
%     \end{array}
%     \right)  \nonumber \\
%     &\times& \sum_{\nk_1, \ldots, \nk_6} \frac{j_{l_1}(k_1 R_D) j_{l_2}(k_2 R_D) j_{l_3}(k_3 R_D) }{(k_1 k_2 k_3 k_4 k_5 k_6 )^{3/2}}  j_{l_1} (k_4 R_D) j_{l_2} (k_5 R_D) j_{l_3} (k_6 R_D) Y_{l_1 m_1}^\star (\hat{k}_1)  \nonumber \\
%     &\times& Y_{l_2 m_2}^\star (\hat{k}_2) Y_{l_3 m_3}^\star (\hat{k}_3)  Y_{l_1 m_4} (\hat{k}_4) Y_{l_2 m_5} (\hat{k}_5) Y_{l_3 m_6} (\hat{k}_6)  T(k_1) T(k_2) T(k_3) T(k_4) T(k_5) T(k_6) \nonumber \\
%     &\times& \overline{X_{\nk_1} X_{\nk_2} X_{\nk_3} X^\star_{\nk_4} X^\star_{\nk_5} X^\star_{\nk_6}}.
% \eea
% where in obtaining  Eq. \eqref{bismodavg} we have  assumed $z_k$ independent of $k$ (see discussion bellow Eq. \eqref{ML6}).
% This assumption is motivated by the results of previous works, in which considering that particular form for the time of collapse,  naturally leads to  a perfect scale-invariant spectrum (i.e. $l(l+1)C(l) =$ constant) . 
As we are assuming that the $X_{\nk}$ follow a highly Gaussian PDF, then we can approximate the 6th momentum distribution by: 

\bea\label{randoms}
\overline{X_{\nk_1} X_{\nk_2} X_{\nk_3} X^\star_{\nk_4} X^\star_{\nk_5} X^\star_{\nk_6}} &=& \nonumber \\
\overline{X_{\nk_1}^{} X_{\nk_2}^{}} \cdot \overline{X_{\nk_3}^{} X_{\nk_4}^{\star}} \cdot \overline{X_{\nk_5}^{\star} X_{\nk_6}^{\star}} &+& \overline{X_{\nk_1}^{} X_{\nk_2}^{}} \cdot \overline{X_{\nk_3}^{} X_{\nk_5}^{\star}} \cdot \overline{X_{\nk_4}^{\star} X_{\nk_6}^{\star}} + \overline{X_{\nk_1}^{} X_{\nk_2}^{}} \cdot \overline{X_{\nk_3}^{} X_{\nk_6}^{\star}} \cdot \overline{X_{\nk_4}^{\star} X_{\nk_5}^{\star}} \nonumber \\
+\overline{X_{\nk_1}^{} X_{\nk_3}^{}} \cdot \overline{X_{\nk_2}^{} X_{\nk_4}^{\star}} \cdot \overline{X_{\nk_5}^{\star} X_{\nk_6}^{\star}} &+& \overline{X_{\nk_1}^{} X_{\nk_3}^{}} \cdot \overline{X_{\nk_2}^{} X_{\nk_5}^{\star}} \cdot \overline{X_{\nk_4}^{\star} X_{\nk_6}^{\star}} + \overline{X_{\nk_1}^{} X_{\nk_3}^{}} \cdot \overline{X_{\nk_2}^{} X_{\nk_6}^{\star}} \cdot \overline{X_{\nk_4}^{\star} X_{\nk_5}^{\star}} \nonumber \\
+\overline{X_{\nk_1}^{} X_{\nk_4}^{\star}} \cdot \overline{X_{\nk_2}^{} X_{\nk_3}^{}} \cdot \overline{X_{\nk_5}^{\star} X_{\nk_6}^{\star}} &+& \overline{X_{\nk_1}^{} X_{\nk_4}^{\star}} \cdot \overline{X_{\nk_2}^{} X_{\nk_5}^{\star}} \cdot \overline{X_{\nk_3}^{} X_{\nk_6}^{\star}} + \overline{X_{\nk_1}^{} X_{\nk_4}^{\star}} \cdot \overline{X_{\nk_2}^{} X_{\nk_6}^{\star}} \cdot \overline{X_{\nk_3}^{} X_{\nk_5}^{\star}} \nonumber \\
+\overline{X_{\nk_1}^{} X_{\nk_5}^{\star}} \cdot \overline{X_{\nk_2}^{} X_{\nk_3}^{}} \cdot \overline{X_{\nk_4}^{\star} X_{\nk_6}^{\star}} &+& \overline{X_{\nk_1}^{} X_{\nk_5}^{\star}} \cdot \overline{X_{\nk_2}^{} X_{\nk_4}^{\star}} \cdot \overline{X_{\nk_3}^{} X_{\nk_6}^{\star}} + \overline{X_{\nk_1}^{} X_{\nk_5}^{\star}} \cdot \overline{X_{\nk_2}^{} X_{\nk_6}^{\star}} \cdot \overline{X_{\nk_3}^{} X_{\nk_4}^{\star}} \nonumber \\
+\overline{X_{\nk_1}^{} X_{\nk_6}^{\star}} \cdot \overline{X_{\nk_2}^{} X_{\nk_3}^{}} \cdot \overline{X_{\nk_4}^{\star} X_{\nk_5}^{\star}} &+& \overline{X_{\nk_1}^{} X_{\nk_6}^{\star}} \cdot \overline{X_{\nk_2}^{} X_{\nk_4}^{\star}} \cdot \overline{X_{\nk_3}^{} X_{\nk_5}^{\star}} + \overline{X_{\nk_1}^{} X_{\nk_6}^{\star}} \cdot \overline{X_{\nk_2}^{} X_{\nk_5}^{\star}} \cdot \overline{X_{\nk_3}^{} X_{\nk_4}^{\star}}. \nonumber \\
\eea
Therefore, substituting Eq. \eqref{randoms} into Eq. \eqref{alm6ptos} yields

\bea
& & \almsix = \nonumber \\
& &\frac{1}{L^6} \sum_{\nk_1, \ldots, \nk_6} g(z_{k_1}) g(z_{k_2}) g(z_{k_3}) g(z_{k_4}) g(z_{k_5}) g(z_{k_6})  F_{l_1 m_1} (\nk_1) F_{l_2 m_2} (\nk_2) F_{l_3 m_3} (\nk_3)    F_{l_1 m_4}^* (\nk_4) F_{l_2 m_5}^* (\nk_5) F_{l_3 m_6}^* (\nk_6) \nonumber\\
&\times& \left(  \overline{X_{\nk_1} X_{\nk_2}}\cdot \overline{X_{\nk_3} X_{\nk_4}^*} \cdot \overline{ X_{\nk_5}^* X_{\nk_6}^*}  + 14\text{ permutations}      \right) \nonumber \\
&=& \frac{1}{L^6} \left( \sum_{\nk_1,\nk_2} g(z_{k_1}) g(z_{k_2}) F_{l_1 m_1} (\nk_1) F_{l_2 m_2} (\nk_2) \overline{X_{\nk_1} X_{\nk_2}}  \right) \left( \sum_{\nk_3,\nk_4} g(z_{k_3}) g(z_{k_4}) F_{l_3 m_3} (\nk_3) F_{l_1 m_4}^* (\nk_4) \overline{X_{\nk_3} X_{\nk_4}^*}  \right) \nonumber \\
&\times& \left( \sum_{\nk_5,\nk_6} g(z_{k_5}) g(z_{k_6}) F_{l_1 m_1}^* (\nk_5) F_{l_2 m_2}^* (\nk_6) \overline{X_{\nk_5}^* X_{\nk_6}^*}  \right) + 14 \textrm{ permutations} \nonumber \\
&=& \overline{a_{l_1 m_1} a_{l_2 m_2}} \cdot \overline{a_{l_3 m_3} a_{l_1 m_4}^*} \cdot \overline{a_{l_2 m_5}^* a_{l_3 m_6}^*}+ 14 \textrm{ permutations,}
\eea
where the last step uses Eq. \eqref{alm2ptos0}. This is an expected result from the beginning because the $a_{lm}$ and the $X_{\nk}$ are linearly related, hence,  the $a_{lm}$ must have the same statistics as the $X_{\nk}$, which we have assumed to be highly Gaussian. Thus, all higher order correlators must either be zero (odd) or products of the two-point correlator (even). Of the 15 permutations, there are 6 of the form $\overline{aa^*} \cdot \overline{aa^*} \cdot \overline{aa^*} $ and 9 of the form $\overline{aa} \cdot \overline{aa^*} \cdot \overline{a^*a^*} $. Let us consider the second case first and its relation with  $\overline{|\mathcal{B}_{l_1 l_2 l_3}|^2}$; e.g. let us analyze the term

\bea
\sum_{m_1, \ldots, m_6} & & \left( \begin{array}{lcr}
       l_1& l_2 & l_3  \\
      m_1 & m_2 & m_3
     \end{array}
     \right)  \left( \begin{array}{lcr}
       l_1& l_2 & l_3  \\
      m_4 & m_5 & m_6
     \end{array}
     \right)  \overline{a_{l_1 m_1} a_{l_2 m_2}} \cdot \overline{a_{l_3 m_3} a_{l_1 m_4}^*} \cdot \overline{a_{l_2 m_5}^* a_{l_3 m_6}^*} \nonumber\\
      &=&  \sum_{m_1, \ldots, m_6}  \left( \begin{array}{lcr}
       l_1& l_2 & l_3  \\
      m_1 & m_2 & m_3
     \end{array}
     \right)  \left( \begin{array}{lcr}
       l_1& l_2 & l_3  \\
      m_4 & m_5 & m_6
     \end{array}
     \right) (-1)^{m_1} D_{l_1} \delta_{l_1,l_2} \delta_{m_1,-m_2} D_{l_3} \delta_{l_3,l_1} \delta_{m_3,m_4} (-1)^{m_5} D_{l_2} \delta_{l_2,l_3} \delta_{m_5,-m_6} \nonumber \\ 
     &=& D_{l_1} D_{l_2} D_{l_3} \delta_{l_1,l_2} \delta_{l_2,l_3} \sum_{m_1, m_3, m_5}  \left( \begin{array}{lcr}
       l_1& l_1 & l_3  \\
      m_1 & -m_1 & m_3
     \end{array}
     \right)  \left( \begin{array}{lcr}
       l_3& l_1 & l_1  \\
      m_3 & m_5 & -m_5
     \end{array}
     \right) (-1)^{m_1+m_5} \nonumber \\
&=& D_{l_1} D_{l_2} D_{l_3} \delta_{l_1,l_2} \delta_{l_2,l_3} \left( \sum_{m_1}   \left( \begin{array}{lcr}
       l_1& l_1 & l_3  \\
      m_1 & -m_1 & 0
     \end{array}
     \right)  (-1)^{m_1}      \right) \left(\sum_{m_5}    \left( \begin{array}{lcr}
       l_1& l_1 & l_3  \\
      m_5 & -m_5 & 0
     \end{array}
     \right)   (-1)^{m_5}     \right) \nonumber\\
     &=& D_{l_1} D_{l_2} D_{l_3} \delta_{l_1,l_2} \delta_{l_2,l_3} (2l_1+1) \delta_{l_3,0},
     \eea
where in the second line we used the relations given in Eq. \eqref{alm2ptosa}; in the fourth line we used the selection rule for Wigner 3-$j$ symbols, and in the last line we used an identity for the sum of the Wigner 3-$j$ symbols. Consequently, since this term is proportional to the monopole ($l=0$), which is unobservable, then it vanishes. The same occurs in all 9 terms. Henceforth, all that remains are the 6 terms that we will consider now. Once again, focusing a typical term, e.g.

\bea
\sum_{m_1, \ldots, m_6} & & \left( \begin{array}{lcr}
       l_1& l_2 & l_3  \\
      m_1 & m_2 & m_3
     \end{array}
     \right)  \left( \begin{array}{lcr}
       l_1& l_2 & l_3  \\
      m_4 & m_5 & m_6
     \end{array}
     \right)  \overline{a_{l_1 m_1} a_{l_1 m_4}^*} \cdot \overline{a_{l_2 m_2} a_{l_2 m_5}^*} \cdot \overline{a_{l_3 m_3} a_{l_3 m_6}^*} \nonumber\\
     &=&  \sum_{m_1, \ldots, m_6}  \left( \begin{array}{lcr}
       l_1& l_2 & l_3  \\
      m_1 & m_2 & m_3
     \end{array}
     \right)  \left( \begin{array}{lcr}
       l_1& l_2 & l_3  \\
      m_4 & m_5 & m_6
     \end{array}
     \right)  D_{l_1}  \delta_{m_1,m_4} D_{l_2}  \delta_{m_2,m_5}  D_{l_3} \delta_{m_3,m_6} \nonumber \\ 
     &=& D_{l_1} D_{l_2} D_{l_3} \left[ \sum_{m_1, m_2, m_3} \left( \begin{array}{lcr}
       l_1& l_2 & l_3  \\
      m_1 & m_2 & m_3
     \end{array}
     \right) \right]^2 \nonumber \\
     &=& D_{l_1} D_{l_2} D_{l_3},
     \eea
where in the second line we used Eq. \eqref{alm2ptosa} and in the last line we used an identity of the Wigner 3-$j$ symbols. The remaining 5 terms are computed in a similar way. The final result is thus

\beq\label{bispecavg}
\overline{|\mathcal{B}_{l_1 l_2 l_3}|^2} = D_{l_1} D_{l_2} D_{l_3} \left[1 + \delta_{l_1, l_2} (-1)^{l_3+2l_1} +\delta_{l_2, l_3} (-1)^{l_1+2l_2}+\delta_{l_3, l_1} (-1)^{l_2+2l_3}   +   2 \delta_{l_1,l_2} \delta_{l_2,l_3} \right],
\eeq% 
with $l=1,2, \ldots$. Consequently, the most likely value for the magnitude of the collapse bispectrum is  $|\mathcal{B}_{l_1 l_2 l_3}|_{\textrm{M.L.}} = \left( \overline{|\mathcal{B}_{l_1 l_2 l_3}|^2} \right)^{1/2}$, i.e.

\bea\label{bispeclm}
|\mathcal{B}_{l_1 l_2 l_3}|_{\textrm{M.L.}} &=& \frac{1}{ \pi^{3/2}} \left( \frac{H}{10 M_P \epsilon^{1/2}} \right)^3  \left| \cos z - \frac{\sin z}{z} \right|^3 \left\{ \frac{ [1+\lambda G(l_1)] [1+\lambda G(l_2)] [1+\lambda G(l_3)]  }{ l_1(l_1 +1) l_2(l_2 +1) l_3 (l_3+1)} \right\}^{1/2} \nonumber \\
&\times& \left[1 + \delta_{l_1, l_2} (-1)^{l_3+2l_1} +\delta_{l_2, l_3} (-1)^{l_1+2l_2}+\delta_{l_3, l_1} (-1)^{l_2+2l_3}   +   2 \delta_{l_1,l_2} \delta_{l_2,l_3} \right]^{1/2},
\eea
where we used the definition of $D_l$ and $g(z)$ given in Eqs. \eqref{Dl} and \eqref{gz} respectively.

% [ D  Arregle  hasta  aca.  ]

Equation \eqref{bispeclm} is the main result of this section. Given the definition %of the collapse bispectrum [see.
of Eq. \eqref{bispecorig},  $l_1,l_2,l_3$ must correspond to the sides of a triangle, otherwise $|\mathcal{B}_{l_1 l_2 l_3}|_{\textrm{M.L.}} =0$ just as in  standard treatments. 

%Furthermore, if such ``triangle'' has different side lengths ($l_1 \neq l_2 \neq l_3$), then $\Delta_{l_1 l_2 l_3}$ vanishes exactly (but not $|\mathcal{B}^{\text{obs}}_{l_1 l_2 l_3}|_{\textrm{M.L.}}$). However, if $l_1,l_2,l_3$ are associated with the sides of an isosceles ($l_i = l_j \neq l_k$) or an equilateral triangle ($l_1 = l_2 = l_3$) the terms appearing in $\Delta_{l_1 l_2 l_3}$, contribute to the collapse bispectrum, which generically does not vanishes (e.g., $|\mathcal{B}^{\text{obs}}_{2 2 2}|_{\textrm{M.L.}}^2 = g(z)^6/[(2\pi)^9 6^2]$).

 However we must emphasize that  $|\mathcal{B}_{l_1 l_2 l_3}|$ is not exactly the 
 same as  the magnitude of the traditional theoretical angle-averaged-bispectrum $|B_{l_1 l_2 l_3}|$, as the latter would correspond to the  ensemble  average as
defined in Eq. \eqref{avg}. In fact, in the conventional approach, one would 
relate an average over an ensemble of universes to  a certain quantum  
three-point function,  which   would   of course  vanish  in the  absence of 
``non-Gaussianities.''  In the   standard  approach argument then   proceeds,  by  identifying   the average over the  ensemble of possible universes  with the angle-averaged-bispectrum $B_{l_1 l_2 l_3}$, which is  
just a  orientation average in  the only  universe  we  have access to. Thus  the   theoretical  quantity is    connected    with    suitable 
averages  over   $m$'s of   the quantity $a_{l_1 m_1} a_{l_2 m_2} a_{l_2 m_3}$ 
measured in our own (single) universe.

Our point is,  thus,  connected  with the  fact  that these series of identifications do not   have a  fully transparent and  clear justification (as  discussed  in detail in  Ref. \cite{susana2013}); and  this, in our view, makes it  very difficult to deal, in a  clear  way,  with   existing and potential delicate issues.  One example of such difficulty   arises when   wanting  to study   issues concerning  say large  deviations from isotropy in  the observed  CMB.  On the other hand,  within the collapse model, we find a prediction for the most likely value of  $|\mathcal{B}_{l_1 l_2 l_3}|_{\textrm{M.L.}}$ which can be related directly with the  actual observational quantity: 

\beq\label{avgbispecobs}
|B_{l_1 l_2 l_3}|_{\textrm{  Actual obs}} \equiv \left| \sum_{m_i} \left( \begin{array}{lcr}
      l_1& l_2 & l_3  \\
     m_1 & m_2 & m_3
    \end{array}
    \right) (a_{l_1 m_1} a_{l_2 m_2} a_{l_2 m_3})_{\textrm{ Actual obs}} \right|,
    \eeq
in a much  more   direct and transparent manner. Note,  however,  that  we  make  a distinction  between the theory's  prediction  for the most likely  value of the  observational quantity $|\mathcal{B}_{l_1 l_2 l_3}|_{\textrm{M.L.}}$  and the  actually  observed  quantity itself  $|B_{l_1 l_2 l_3}|_{\textrm{Actual obs}}$.  This is  exactly   analogous to the distinction  beteen a theory  prediction of the most likely value of the total displacement in a  random  walk, and the  actual  displacement observed  in a  single    realization of   said walk.

At this point it is also convenient to clarify certain aspects within our approach. Some readers may claim that our expression for the observed bispectrum is equivalent to the cosmic variance of the bispectrum, within the standard approach, in the ``weak non-Gaussian'' limit. In other words, the fact that we used the approximation in Eq. \eqref{randoms}, corresponding to a highly Gaussian PDF for the random parameters of the collapse, could lead some people to state that what we are computing is actually the variance of the traditional bispectrum, Eq. \eqref{avg}, when $B_{l_1,l_2,l_3}=0$. This is partially correct and a few remarks are in order. 

If $\lambda=0$, which means that the PDF of the random collapse parameters is exactly Gaussian, and if $l_1+l_2+l_3$ is even, then our expression for $|\mathcal{B}_{l_1 l_2 l_3}|_{\textrm{M.L.}}$  coincides   exactly  with the  traditional  estimation of  the  variance of the  bispectrum in the Gaussian limit \cite{Luo1994,Gangui1999,Gangui2000}, i.e. $\textrm{Var}[B_{l_1 l_2 l_3}] \propto C_{l_1} C_{l_2} C_{l_3}$, with $C_l$ the angular power spectrum which is of the form $C_l \propto 1/l(l+1)$ in the range $l \leq 20$. On the other hand, if $\lambda \neq 0$, which physically means that the collapse of the wave function correlates different modes, then our expression \eqref{bispeclm}, contains a function $G(l)$ that  is absent in the traditional approach. Furthermore, as we mentioned previously, in our approach there is in principle no restriction on  $l_1+l_2+l_3$  (which  in the   standard  approach  must be  even  a  and represents an  extra  requirement for the nonvanishing of the bispectrum) as we are not assuming that the ensemble average of the bispectrum coincides with the observed one, this is, we do not basin  our analysis  in anything analogous to the   standard  claim that the 3-point correlation function of the temperature anisotropies must be rotational and translational invariant. This    aspect of  our analysis,  represent  a   difference   in  the  predictions  that  can   be  confronted with   observations.   In the next section, we will  extend  this and other discussions about the relation between the statistical aspects of the traditional bispectrum and the collapse bispectrum.

\section{Main differences between the standard and the collapse approach regarding the primordial bispectrum}\label{diferencias}

\subsection{The standard approach to the CMB bispectrum}

We begin this section by giving a rather brief review of the conventional approach to the  study  of the primordial bispectrum, its amplitude and the usual arguments given to relate it  with possible non-Gaussian features in the perturbations; extended reviews can be found in Refs. \cite{bartolo2004,komatsu2003,Yadav,Liguori,Komatsu2001}.

According  to  the standard approach, if  $\Psi(\x,\eta)$ is  taken to  be  characterized  by a  Gaussian distribution all its statistical properties are codified  in the two-point correlation function. Otherwise, one needs to consider higher order correlation functions, e.g. the three-point correlation function $\overline{\Psi(\x,\eta) \Psi(\vec{y},\eta) \Psi (\vec{z},\eta)}$. The Fourier transform is commonly referred as the bispectrum, defined by

\beq\label{3ptospsi}
\overline{\Psi_{\nk_1} \Psi_{\nk_2} \Psi_{\nk_2}} \equiv (2\pi)^3 \delta (\nk_1 + \nk_2 + \nk_3) B_\Psi (k_1,k_2,k_3). 
\eeq
The Dirac delta appearing in Eq. \eqref{3ptospsi}  is said to  indicate that the ensemble average $\overline{\Psi (\vec{x},\eta) \Psi(\vec{y},\eta) \Psi(\vec{z},\eta)}$  is invariant under spatial  translations; in addition, the dependence only on the magnitudes $k$, appearing  in the function $B_\Psi (k_1,k_2,k_3)$, is  tied   to  the rotational  invariance of such ensemble. The Dirac delta  constrains  the three modes  involved; this is, the modes must satisfy $\nk_1 + \nk_2 + \nk_3 =0$, which is known as the triangle condition.  Thus,   according to  the direct relation of the Newtonian potential with  the   source of the  observed  anisotropies,  the assumption  about  the rotational and translational invariance of the ensemble  and   its  impact on $\overline{\Psi (\vec{x},\eta) \Psi(\vec{y}, \eta) \Psi(\vec{z},\eta)}$ should  be reflected  in  the invariance of the ensemble average of the temperature fluctuations $ \overline{\frac{\delta T}{T_0} (\theta_1, \varphi_1) \frac{\delta T}{T_0} {(\theta_2,\varphi_2)} \frac{\delta T}{T_0} (\theta_3, \varphi_3)}$.  However,  we  should   once  again  emphasize    that strictly speaking the  discussion  above  and, thus, the  average  indicated by  the overline in the latter expressions,  refers to an   average over an ensemble of universes and not averages over orientations in one universe.  

One of the first (and most popular ways) to parameterize the traditional non-Gaussianity phenomenologically is via   the introduction of a non-linear correction to the linear Gaussian curvature perturbation
\cite{salopek1990,salopek1991},

\beq\label{psiloc}
\Psi (\x,\eta) =  \Psi_g (\x,\eta) + f_{\textrm{NL}}^\textrm{loc} [ \Psi_g^2 (\x,\eta) - \overline{\Psi_g^2 (\x,\eta)} ],
\eeq
where $\Psi_g (\x,\eta)$ denotes a linear Gaussian part of the perturbation and $\overline{\Psi_g^2 (\x,\eta)}$ is the variance of the of the Gaussian part.\footnote{Let us recall that the
variance $\overline{\Psi^2 (\x,\eta)} = \int_0^\infty dk k^2 P_\Psi(k,\eta)$ diverges logarithmically  for a power spectrum such that $P_\Psi (k,\eta) \propto k^{-3}$ unless one introduce an 
\emph{ad hoc} cutoff for $k$. For a detailed discussion of this  and other related issues see Ref. \cite{susana2013} and Appendix B of the same Ref. } 
The parameter $f_{\textrm{NL}}^{\textrm{loc}}$ is called the ``local non-linear coupling parameter" and determines the ``strength" of the primordial non-Gaussianity. This parametrization 
of non-Gaussianity is local in real space and therefore is called ``local non-Gaussianity.''  Using Eqs. \eqref{3ptospsi} and \eqref{psiloc}, the bispectrum of local non-Gaussianity may be 
derived:

\bea\label{bispeck}
B_\Psi (\nk_1,\nk_2,\nk_3) &=& 2 f_{\textrm{NL}}^\textrm{loc} [ P_\Psi (\nk_1) P_\Psi (\nk_2) + P_\Psi (\nk_2) P_\Psi (\nk_3) + P_\Psi (\nk_3) P_\Psi (\nk_1) ] \nonumber \\
&=& 2 f_{\textrm{NL}}^\textrm{loc} \frac{A^2_{\Psi}}{(k_1 k_2 k_3)^2} \left(\frac{k_1^2}{k_2k_3} + \frac{k_2^2}{k_1k_3} + \frac{k_3^2}{k_1k_2} \right),
\eea
with $P_\Psi (\nk)$ the power spectrum of the Newtonian potential  defined as $\overline{\Psi (\nk) \Psi (\nk')} \equiv (2\pi)^3 \delta (\nk + \nk') P_\Psi (k)$ and it is assumed to be of the form $P_\Psi (\nk) = A_\Psi k^{-3}$, where $A_\Psi$ denotes the amplitude of the power spectrum. 

Expressing the coefficients $a_{lm}$ in terms of $\Psi_{\nk}$ as in Eq. \eqref{alm2} and using Eq. \eqref{3ptospsi}, one can proceed to compute the CMB bispectrum, defined in Eq. \eqref{bispec0}, this is

\bea\label{fullbis}
B^{l_1 l_2 l_3}_{m_1 m_2 m_3} &=& \overline{a_{l_1 m_1} a_{l_2 m_2} a_{l_3 m_3}} \nonumber \\
&=&  \bigg( \frac{4\pi}{3} \bigg)^3 i^{l_1 + l_2 + l_3} \int \frac{d^3 k_1}{(2\pi)^3} \frac{d^3 k_2}{(2\pi)^3} \frac{d^3 k_3}{(2\pi)^3}  j_{l_1} (k_1 R_D) j_{l_2} (k_2 R_D) j_{l_3} (k_3 R_D)   \nonumber \\ 
&\times& T(k_1) T(k_2) T(k_3) \overline{\Psi_{\nk_1} \Psi_{\nk_2} \Psi_{\nk_3} } Y_{l_1 m_1} (\hat{k_1} ) Y_{l_2 m_2} (\hat{k_2} ) Y_{l_3 m_3} (\hat{k_3} ) \nonumber \\
&=& \bigg( \frac{2}{\pi} \bigg)^3  \int dk_1 dk_2 dk_3 \textrm{ } (k_1 k_2 k_3)^2  T(k_1) T(k_2) T(k_3) j_{l_1} (k_1 R_D) j_{l_2} (k_2 R_D) j_{l_3} (k_3 R_D) \nonumber \\
&\times&  B_\Psi (k_1,k_2,k_3) \int_0^\infty dx \textrm{ } x^2 j_{l_1} (k_1 x) j_{l_2} (k_2 x) j_{l_3} (k_3 x)  \int d \Omega_{\hat{x}} Y_{l_1 m_1} (\hat{x} ) Y_{l_2 m_2} (\hat{x} ) Y_{l_3 m_3} (\hat{x} ), \nonumber\\
\eea
where in the last line, one performs the integral over the directions of the three vectors $k_i$ and use the exponential integral form for the delta function that appears in the bispectrum definition \eqref{3ptospsi}. The integral over the angular part of $\vec x$, in the last line, is known as the Gaunt integral, i.e.,

\begin{eqnarray}\label{gaunt}
\mathcal{G}_{l_1 l_2 l_3}^{m_1 m_2 m_3} &\equiv& \int d \Omega_{\hat{x}} Y_{l_1 m_1} (\hat{x} ) Y_{l_2 m_2} (\hat{x} ) Y_{l_3 m_3} (\hat{x} ) \nonumber \\
&=& \sqrt{\frac{(2l_1+1)(2l_2+1)(2l_3+1) }{ 4\pi } } \left( \begin{array}{lcr}
      l_1& l_2 & l_3  \\
     0 & 0 & 0
    \end{array}
    \right) \left( \begin{array}{lcr}
      l_1& l_2 & l_3  \\
     m_1 & m_2 & m_3
    \end{array}
    \right).
\end{eqnarray}
The fact that the bispectrum $B_{m_1 m_2 m_3}^{l_1 l_2 l_3}$  turns  out to be   proportional  of the Gaunt integral, $\mathcal{G}_{l_1 l_2 l_3}^{m_1 m_2 m_3}$, implies that the values of $l,m$, corresponding to the non-vanishing  components  of  the  bispectrum,  must satisfy ``the triangle conditions'' and also reflect   the fact that  in the definition of the bispectrum there  is  a Dirac delta [Eq. \eqref{3ptospsi}] which  guarantees the  rotational invariance for the ensemble average $\overline{\Psi_{\nk_1} \Psi_{\nk_2} \Psi_{\nk_3}}$. 

One can find a close form for the bispectrum by choosing to work in the Sach-Wolfe approximation, where the transfer function $T(k)=1$; thus, substituting Eq. \eqref{bispeck} into Eq. \eqref{fullbis} and evaluating  the remaining integrals one obtains (See Ref. \cite{fergusson2009}):

\bea\label{bispecloc44}
B^{l_1 l_2 l_3}_{m_1 m_2 m_3} &=& \mathcal{G}_{l_1 l_2 l_3}^{m_1 m_2 m_3} f_{\textrm{NL}}^{\textrm{loc}} \left( \frac{2 A_{\Psi}^2}{27 \pi^2} \right) \bigg( \frac{1}{l_1(l_1+1)l_2(l_2+1)} + 
\frac{1}{l_2(l_2+1)l_3(l_3+1)} \nonumber \\
&+& \frac{1}{l_1(l_1+1)l_3(l_3+1)} \bigg).
\eea
Finally, using the definition of the angle-averaged-bispectrum [Eq. \eqref{avg}], one has

\bea\label{bispecloc0}
B_{l_1 l_2 l_3} &=& \sqrt{\frac{(2l_1+1)(2l_2+1)(2l_3+1)}{4\pi}} \left( \begin{array}{lcr}
      l_1& l_2 & l_3  \\
     0 & 0 & 0
    \end{array}
    \right) f_{\textrm{NL}}^{\textrm{loc}} \left( \frac{2 A_{\Psi}^2}{27 \pi^2} \right) \nonumber \\
    &\times&  \left( \frac{1}{l_1(l_1+1)l_2(l_2+1)} + 
\frac{1}{l_2(l_2+1)l_3(l_3+1)} + \frac{1}{l_1(l_1+1)l_3(l_3+1)} \right).
\eea
When working   within  the standard scenario for slow-roll inflation (and thus a ``nearly" scale-invariant spectrum), Maldacena found that the estimate  for the amplitude of non-Gaussianities of  the local form, is  $f_{\textrm{NL}}^{\textrm{loc}} \simeq \epsilon$  \cite{maldacena2002} (this is assumed to be in the limit when $k_1 \ll k_2 \simeq k_3$, i.e. in the so-called ``squeezed''  configuration). As seen from the previous discussion, in the conventional approach, the amplitude of the primordial bispectrum and the non-Gaussian statistics for the curvature  perturbation are intrinsically related.

\subsection{Comparing the magnitude of the collapse and the traditional bispectrum}

The magnitude of $|B_{l_1 l_2 l_3}|$ is obtained from Eq. \eqref{bispecloc0}

\bea\label{bispecloc}
|B_{l_1 l_2 l_3}| &=& \sqrt{\frac{(2l_1+1)(2l_2+1)(2l_3+1)}{4\pi}} \left| \left( \begin{array}{lcr}
      l_1& l_2 & l_3  \\
     0 & 0 & 0
    \end{array}
    \right) \right| |f_{\textrm{NL}}^{\textrm{loc}}| \left( \frac{2 H^4}{27 \pi^2 M_P^4 \epsilon^2} \right) \nonumber \\
    &\times&  \left( \frac{1}{l_1(l_1+1)l_2(l_2+1)} + \frac{1}{l_2(l_2+1)l_3(l_3+1)} + \frac{1}{l_1(l_1+1)l_3(l_3+1)} \right);
\eea
where  we have used the   estimate  for the amplitude of the power spectrum for a single scalar field in the slow-roll scenario,  given by $A_\Psi \simeq H^2/(M_P^2 \epsilon)$. 

On the other hand, the magnitude of the collapse biscpectrum is given by Eq. \eqref{bispeclm}, which we will write again

\bea\label{biscolapso}
|\mathcal{B}_{l_1 l_2 l_3}|_{\textrm{M.L.}} &=& \frac{1}{ \pi^{3/2}} \left( \frac{H}{10 M_P \epsilon^{1/2}} \right)^3  \left| \cos z - \frac{\sin z}{z} \right|^3 \left\{ \frac{ [1+\lambda G(l_1)] [1+\lambda G(l_2)] [1+\lambda G(l_3)]  }{ l_1(l_1 +1) l_2(l_2 +1) l_3 (l_3+1)} \right\}^{1/2} \nonumber \\
&\times& \left[1 + \delta_{l_1, l_2} (-1)^{l_3+2l_1} +\delta_{l_2, l_3} (-1)^{l_1+2l_2}+\delta_{l_3, l_1} (-1)^{l_2+2l_3}   +   2 \delta_{l_1,l_2} \delta_{l_2,l_3} \right]^{1/2}.
\eea

Therefore we have two distinct theoretical predictions for the actual observed bispectrum $|B_{l_1 l_2 l_3}|_{\textrm{Actual obs}}$ [see Def. \eqref{avgbispecobs}]. In the standard slow-roll inflationary scenario the prediction is given by $|{B}_{l_1 l_2 l_3}|$ Eq. \eqref{bispecloc}; meanwhile, by considering the collapse hypothesis the prediction is $|\mathcal{B}_{l_1 l_2 l_3}|_{\textrm{M.L.}}$ Eq. \eqref{biscolapso}.

The first  and most important difference is the fact that Eq.  \eqref{bispecloc}   vanishes  unless   $f_{\textrm{NL}}^{\textrm{loc}} \not=0$,  while   no such ``primordial  non-Gaussianity'' is  required for a non-vanishing value of Eq. \eqref{biscolapso}.

%The second difference is that in the traditional bispectrum, the values of $l_i,m_i$ with $i=1,2,3$ are constrained by the Gaunt integral, i.e. they must satisfy the triangle conditions otherwise  $|B^{l_1 l_2 l_3}_{m_1 m_2 m_3}| = 0$; in fact the total dependence on $m_i$ is contained in the Gaunt integral. In  contrast, the collapse bispectrum does not impose any restriction for the possible values of $l_i,m_i$.  

The second difference is that the shape of the bispectrum, i.e. its dependence on $l$, is not the same; $|\mathcal{B}^{\text{obs}}_{l_1 l_2 l_3}|_{\textrm{M.L.}}$ scales roughly as $\sim[(1+\lambda G(l))/l(l+1)]^{3/2}$ while $|{B}_{l_1 l_2 l_3}|$ as $\sim (2l+1)^{3/2}/[l^3(l+1)^2]$. Unfortunately, the  existing  analysis  of the observational data do not focus the exact shape of the biscpectrum, but rather  on a generic  measure  of   its amplitude. However, the reported observational amplitude of the bispectrum, which in the standard picture of slow-roll inflation  corresponds to  the non-linear parameter $f_{\textrm{NL}}^{\textrm{loc}}$, depends  on the expected  shape  that emerges  from  the theoretical  estimates  of bispectrum \cite{Yadav,Liguori}. In other words,   in the standard approach,   in order to obtain an estimate for the amplitude of the bispectrum from the observational 
data, one requires a theoretical motivated shape for the bispectrum. The  observational data and theory are  strongly   interdependent. 
In  this  way,   and relying on such theoretical  considerations,  the latest results 
from \emph{Planck} mission \cite{planckng}  lead to an estimate for the amplitudes of the bispectrum for the local, equilateral, and orthogonal 
models   given by 
$f_{\textrm{NL}}^{\textrm{loc}} = 2.7 \pm 5.8, f_{\textrm{NL}}^{\textrm{equil}} = -42 \pm 75 $, and $f_{\textrm{NL}}^{\textrm{ortho}} = -25 \pm 39 $ (68\% CL statistical). On the other hand, none of the previously considered  shapes, namely the local, equilateral or orthogonal, correspond to the one given by the collapse bispectrum.

% 
% It is also interesting to note that, in the traditional framework, there is a well known result \cite{creminelli} when considering  single-field inflation named the ``consistency relation.'' This is, the value predicted for the local non-linear parameter is given by $f_{\textrm{NL}}^{\textrm{loc}} \simeq n_s-1$, with $n_s$ the ``spectral index'' of the primordial scalar fluctuations. This prediction is obtained by 
% assuming a single scalar field and  no other assumptions  (within the  standard approach); in particular, it is independent of: the form of the potential, the form of the kinetic term and the initial vacuum state.   It is clear 
% that for a perfect scale-invariant spectrum, that is when $n_s=1$,  the standard prediction is $f_{\textrm{NL}}^{\textrm{loc}}=0$, which  according  to standard  estimates implies the prediction $|B_{l_1 l_2 l_3}| = 0$. In contrast, within the collapse picture,   there is  no  such  close connection between  spectral index  and the collapse bispectrum. Thus, as we proved at the end of Sec. \ref{angularspectrum}, assuming a perfect scale-independent spectrum  does not imply that the  predicted  value should be  $|\mathcal{B}^{\text{obs}}_{l_1 l_2 l_3}|_{\textrm{M.L.}} = 0$; in some sense we have given a counter-example for the ``consistency relation.'' 

\subsection{Comparison with the variance of the traditional bispectrum}

Furthermore, one can compute the cosmic variance of the traditional bispectrum, in the case of weak non-Gaussianity. This is, in the Gaussian limit, the variance of the standard bispectrum is \cite{Luo1994,Gangui1999,Gangui2000}

\bea\label{covarbispec}
\overline{|B_{l_1 l_2 l_3}|^2} &=& C_{l_1} C_{l_2} C_{l_3} \left[ 1 + \delta_{l_1,l_2} + \delta_{l_2,l_3} + \delta_{l_3,l_1} + 2 \delta_{l_1,l_2} \delta_{l_2,l_3} \right] \nonumber \\
&\propto& \frac{ 1 }{ l_1(l_1 +1) l_2(l_2 +1) l_3 (l_3+1)}  \left[ 1 + \delta_{l_1,l_2} + \delta_{l_2,l_3} + \delta_{l_3,l_1} + 2 \delta_{l_1,l_2} \delta_{l_2,l_3} \right].
\eea
Meanwhile rewriting Eq. \eqref{bispecavg} yields

\bea\label{bispecavg2}
|\mathcal{B}_{l_1 l_2 l_3}|^2_{\textrm{M.L.}} &=& D_{l_1} D_{l_2} D_{l_3} \left[1 + \delta_{l_1, l_2} (-1)^{l_3+2l_1} +\delta_{l_2, l_3} (-1)^{l_1+2l_2}+\delta_{l_3, l_1} (-1)^{l_2+2l_3}   +   2 \delta_{l_1,l_2} \delta_{l_2,l_3} \right] \nonumber \\
&\propto& \frac{ [1+\lambda G(l_1)] [1+\lambda G(l_2)] [1+\lambda G(l_3)]  }{ l_1(l_1 +1) l_2(l_2 +1) l_3 (l_3+1)} [1 + \delta_{l_1, l_2} (-1)^{l_3+2l_1} +\delta_{l_2, l_3} (-1)^{l_1+2l_2}+\delta_{l_3, l_1} (-1)^{l_2+2l_3} \nonumber \\  
&+&   2 \delta_{l_1,l_2} \delta_{l_2,l_3} ].
\eea

Comparing  Eqs. \eqref{covarbispec} and \eqref{bispecavg2}, we note some similarities but also fundamental differences: First, if $l_1\neq l_2 \neq l_3$ and the sum $l_1+l_2+l_3=$ odd, our theoretical prediction is $|\mathcal{B}_{l_1 l_2 l_3}|^2_{\textrm{M.L.}} = D_{l_1} D_{l_2} D_{l_3} $, meanwhile in the standard approach the theoretical prediction would be $ \overline{|B_{l_1 l_2 l_3}|^2} = 0$. This is because, the traditional approach makes the additional assumption that the 3-point correlation function must be invariant under spatial translations and rotations; in turn, in our approach, since we do not identify theoretical averages with observational quantities, the selection rule  $l_1+l_2+l_3=$ even, does not apply. Moreover, if $l_1+l_2+l_3=$ even and $\lambda=0$, then the variance of the traditional bispectrum and our prediction matches exactly, this is, $\overline{|B_{l_1 l_2 l_3}|^2} = |\mathcal{B}_{l_1 l_2 l_3}|^2_{\textrm{M.L.}} \propto C_{l_1} C_{l_2} C_{l_3}$. Note, however, that in general $\lambda \neq 0$ since this parameter characterize the correlation between modes that is induced by the collapse of the inflaton's wave function. Therefore, generically $\overline{|B_{l_1 l_2 l_3}|^2} \neq |\mathcal{B}_{l_1 l_2 l_3}|^2_{\textrm{M.L.}}$; moreover, there is the issue of  the parity of $l_1+l_2+l_3$,   mentioned above ,  all   of which 
%depends on the statistical aspects one is considering, 
lead to  the  difference in the  estimations of the two approaches reflected in the two  expressions   Eqs. \eqref{covarbispec} and \eqref{bispecavg2}.

We  must of course  acknowledge  that adopting a pragmatic point of view would lead one to argue that within the standard approach, and in the absence of non-gaussianities, the  expected bispectrum would be zero up to the  cosmic variance that can be seen in the observational data. In such case, our prediction, might  be reinterpreted as an ``improved estimate'' of the traditional cosmic variance associated to the bispectrum, and in fact, the effects for introducing the collapse of the wave function, encoded in the function $G(l)$, can be sought within the usual bispectrum cosmic variance.

However, from the conceptual point of view, there are several differences between ours and the standard approach.
Within  the  collapse hypothesis perspective, the observed $|B_{l_1 l_2 l_3}|_{\textrm{Actual obs}}$  is   clearly  identified  as  corresponding  to just one particular realization of a random quantum process (the self-induced collapse of the wave function). Since we do not have access to other realizations (i.e. we do not have observational access to other universes) we cannot   say anything conclusive as to  whether the underlying PDF is Gaussian or not. In other words, by measuring a non-vanishing $|B_{l_1 l_2 l_3}|_{\textrm{Actual obs}}$ in our own universe, which corresponds to a single realization of the physical process,   cannot be taken as  indication  that the ensemble average $\overline{\Psi_{\nk_1} \Psi_{\nk_2} \Psi_{\nk_3}}$, is also non-vanishing and  would  consequently  proving non-Gaussianity statistics for the ensemble.   In fact,  just as   we do not  expect the   actual  value  of   the one  available  realization   of    $a_{lm}$   (for a fixed  value of $l$ and $m$) to  vanish  identically, even if  somehow  the  ensemble  average of such quantity  (something  to which  we  would   have  no  access  even if an  ensemble of universes  did  exist)  vanishes,  we  should not expect the  single realization of a   bispectrum, corresponding to our   observations   of the universe, to  vanish  identically,  even if the  average  value  over  an ensemble of universes would  vanish.

\section{Conclusions}\label{concl}

We  have  presented  a detailed   discussion on   the  manner, in which the   study  of   essential  statistical  features  on the   CMB  spectrum,  must  be    studied  in the context of the collapse  models.  We  have  shown  the  important  differences that   arise between the analysis in this  approach  and those tied  to the standard   approaches. We  have  seen, for instance, that  the collapse models  lead to  explicit expressions  for the quantities  that are    rather directly  observable,  the  $ a_{lm}  $  [see Eq. \eqref{almrandom}], which have no  counterpart in the  usual  analyses.   Among other  advantages,  the  expression for the coefficients $a_{lm}$ [Eq. \eqref{almrandom}]   exhibits   directly the source of the randomness involved, aspects that in the standard approach  can  only be  discussed  heuristically (simply because    there,  the random  variables are not clearly identified  and  named  as in this approach\footnote{ Researchers  involved in direct comparison  of  actual   empirical    data  with   theoretical estimates apparently make use of   similar random parameters  in performing  the simulations   used in the process of  confrontation of theory and  observations \cite{Arthur}}).  We   have seen that this leads to  some  differences  in the   analysis of  the higher   order  statistical features,  such as  the angular power spectrum [Eq. \eqref{clML}] and the  bispectrum [Eq. \eqref{bispecavg}].  Within  the standard approach, the bispectrum is usually associated  with non-Gaussianities, meanwhile in our approach, it is  a  characterization of the limited  observational data    associated  with   the single realization of an explicitly   described   random process. On the other hand, the  usage of the bispectrum to  infer statistical aspects of an hypothetical  ensemble  of   universes,  characterized by a  multiplicity of  realizations  of  primordial perturbations,  can be  seen as  a   involving   rather   delicate  extrapolations.

We have  seen that    in the  collapse  scheme, one is  lead to   expect   generically  a   non-vanishing     bispectrum,   even if  there are no primordial non-Gaussianities (i.e.  there  are no assumptions regarding  a  non-vanishing  quantum-tree  point function,  and in the statistical  analysis   of  Sec. \ref{seccolapsobispec}  all  statistical  correlations  are  encoded  in the   two  variable correlation functions   of the   fiducial   ensemble of universes,   which are taken  as standard).  On the other hand, within the traditional approach, if one assumes that the primordial curvature perturbation is Gaussian, then the bispectrum must be zero up to its cosmic variance Eq. \eqref{covarbispec}. Furthermore, after comparing our prediction for the observed bispectrum and the traditional variance associated to the bispectrum, we found that in general these two expressions are not equal (although for certain particular cases they do coincide). In fact, some of   the differences arise directly from  the  assumptions  regarding   the self-induced collapse of the wave function: First, there is the issue of  the parity of the bispectrum, in the traditional approach $l_1+l_2+l_3$ must be even,  for  a non-vanishing  value of  the bispectrum  in order to  be  in agreement with a statistical homogeneity and isotropy of the ensemble of  temperature anisotropies. In our model, there is no constraint on the parity of $l_1+l_2+l_3$ as  we do not associate  in any  direct way  the average over an ensemble of universes with the actual observed value of the bispectrum, which corresponds to just a particular element of the hypothetical ensemble. Second, the collapse of the wave function induces  small correlations between different modes characterizing the inflaton (where the ``smallness,'' is parameterized by $\lambda$), which in turn, imprints a particular signature in both, the angular spectrum (see Fig. 2) and the bispectrum that is encoded in the function $G(l)$ [see Eqs. \eqref{Gl} and \eqref{bispecavg2}]. Therefore, even if one takes a pragmatical point of view, and argue that what we have actually computed is  ``just'' the variance of the bispectrum, for the case when the primordial perturbations are Gaussian, our  result   differs  in   various mathematical   aspects  from that obtained in the standard approach. These differences  (mainly those  associated  with the parity and the additional function $G(l)$)   indicate that  our prediction must  at least    be  viewed as  a  novel one  and    we  think it constitutes an improved estimate of the  bispectrum cosmic variance. This   makes the analysis  highly    predictive, and  by the  same token,   highly  susceptible  to falsification by observational data, i.e. by using the data analysis, say from Planck collaboration, one could search for a signal that matches our prediction; in particular, the function $G(l)$ mostly affects the lowest multipole angular scales $l \leq 20$ and then it decreases its effect on large multipoles in a nearly exponential fashion (see Figs. 1,2).

% 
%  One  might  be  tempted to   use  the   estimates   for the  bispectrum  amplitude obtained in the analysis of the  data  for the other models,  but as   explained in Sec. \ref{comparisons},  the fact that   the    functional  form   of the expected  bispectrum is  very different in our model   and the  models  that have been used as    for the   comparison  with observations,  invalidates  from the start  that    program.  One can   confirm this fact simply  by  noting the   sharp differences in the estimates  of $f_{\textrm{NL}}$ that are  extracted  from  the same data  for  the various  models  mentioned above.  However, the data  are    in principle available and,   thus,  testing the prediction of the collapse  models   seems    to  be   well within  reach.
 
%%%%%%%%
  Nevertheless,  before even proceeding to  do this,  one would  need  to  reevaluate   the predicted angular spectrum and bispectrum  by taking into account the  effect of the transfer functions that we have   ignored  in performing the calculation leading to    the expressions \eqref{clML} and  \eqref{bispeclm}. That is,  in such calculation, we  replaced the  transfer functions  by   the number $1$   so  that the    integrals could be evaluated in closed   form. Thus, the  actual  comparison   of the prediction of this model  with data   will require  the    reintroduction of the  transfer functions in the evaluation.   Carrying this   out would
  involve  numerical    calculations  in order to perform the desired integrals  that will  result in the  specific   form of the     $\mathcal{B}_{l_1 l_2 l_3}$, which will be  suitable for comparison  with 
  observations.  We  plan to  carry this  analysis in the  near future and   to  obtain the data to  contrast  with the model's   predictions. We  would view  a  reasonable match   between    observations  and  the model as a   strong  indication  we  are on the right track.

  We   end  up  by noting that  there  are  a couple of aspects of our analysis  that   have  suggestive  counterparts in  observations that have from time to time   caught the attention of  some researchers  in the field:
  
  \begin{enumerate}
   \item  There  is  a claim of an  abnormal   deficit in power  in the   spectrum  at   low   values of  $l$'s    \cite{copi,planck2013}  which  we think can be 
    naturally    explained  within  our approach  by a  negative    value of $\lambda $ in   equation  \eqref{clML}.
    
    \item There has   been  some   attention given to  an anomalous   high  degree of  alignment   occurring in the orientation of the low  $l$  multipoles (sometimes called  the ``axis  of  evil'' \cite{evilaxis})  and  it  seems   again that  a  high  correlation  between the  modes  $\vec k$ and  $ 2\vec k$   that  we have  explained is   a  natural   result  of    the collapse  mechanism as  viewed  within the SSC formal  analysis,  would   have the  basic    features   required  to account for that.   
 \end{enumerate}

   These  and  other  points  would  of course  deserve further inquiry   that falls  outside the   scope of the present manuscript.  
  
%   
%   We  would certainly  not  expect  a complete and precise   agreement, simply due to the fact  that,  
%    our model,   allows  us  to obtain only a  most likely  value  for the quantities controlled  by the random numbers associated  with the  collapse processes.  

%However,   the fact that   the 
%   scheme  for  evaluating the  most likely value of the bispectrum is   essentially  the    same  scheme  we  used too evaluate the  spectrum, as well as the  most likely  values of the  $C_{l}$ [see  Eq. \eqref{clML}],
%   and that we found  a   very good  match  between theory and observations,  would make  it very difficult to understand  such agreement  in one case and  any  strong departure in the  second.
 
%   
%   Finally,  in  Sec. \ref{nuevacantidad}, we  have
%  proposed   a new  quantity   for use in the study  of  higher order   statistical   features  of the CMB  data: $\mathcal{F}_{l_1 l_2 l_3}$ [see Eq. \eqref{Fluc}].
%   The  estimate for this     quantity   in  any  of the standard  scenarios  would  vanish, independently of   the  functional form of the  bispectrum,  simply  because
% the  standard  schemes  offer no mathematical  characterization of the   randomness, and  that   should be a key  aspect of  any  description of   something like the   distribution of the   seeds  of cosmic  structure.  In the   collapse  scheme,  we have  a  specific   expression for the   most likely value of this  quantity. Therefore,   a  comparison with the observation would  be  a nice  empirical test of the ideas  tied to our proposals.

\acknowledgments

We wish to thank Prof. Matt Visser and Prof. Arthur Kosowsky for useful suggestions that improved the presentation of the results obtained in this paper. The work of GL is supported by  postdoctoral grant from Consejo Nacional de Investigaciones Cient\'{\i}ficas y T\'{e}cnicas, Argentina. The work  of DS was supported in part by the CONACYT-M\'exico Grant No. 101712 and UNAM-PAPIIT Grant No. IN107412.

%%%%%%%%%%%%%%%%%%%%%%%%%%%%%%

\end{document}